\documentclass[aps,prl,nofootinbib,preprintnumbers,amsmath,amssymb,twocolumn,superscriptaddress,floatfix]{revtex4-1}

\usepackage{epsfig}
\usepackage{array}
\usepackage{epstopdf}
\usepackage{bm}
\usepackage{cancel}
\usepackage{xcolor}
\usepackage{graphicx}
\usepackage{mathrsfs}
\usepackage{slashed}
\usepackage{textcomp}
\usepackage{times}
\usepackage{float}
\usepackage{url}
\usepackage{cancel}
\usepackage{textcomp}
\usepackage{amsmath, amssymb, amsfonts, latexsym, epsfig}
\usepackage{mathrsfs}
\usepackage{hyperref}
\usepackage{mathtools}
\definecolor{linkblue}{rgb}{0.0, 0.0, 0.5}
\hypersetup{
    colorlinks=true,
    linkcolor=linkblue,    
    urlcolor=linkblue,
    citecolor=linkblue
    }

\def\beq{\begin{equation}}
\def\eeq{\end{equation}}
\def\bea{\begin{eqnarray}}
\def\eea{\end{eqnarray}}

\def\lsim{\mathrel{\rlap{\lower4pt\hbox{\hskip1pt$\sim$}}
     \raise1pt\hbox{$<$}}}    
\def\gsim{\mathrel{\rlap{\lower4pt\hbox{\hskip1pt$\sim$}}
     \raise1pt\hbox{$>$}}}

\begin{document}
\title{Leptogenesis with dynamical couplings}
\preprint{CETUP-2023-012}

\author{Peisi Huang}
\email{peisi.huang@unl.edu}
\affiliation{Department of Physics and Astronomy$,$ University of Nebraska$,$ Lincoln$,$ NE 68588$,$ USA}
\author{Tao Xu}
\email{tao.xu@ou.edu}
\affiliation{Homer L. Dodge Department of Physics and Astronomy$,$ University of Oklahoma$,$ Norman$,$ OK 73019$,$ USA}

\begin{abstract}
We propose a novel leptogenesis mechanism with a temperature-dependent coupling between the right-handed neutrino and Standard Model particles. This coupling experiences suppression at high temperatures and becomes sizable when the lepton asymmetry washout processes are Boltzmann-suppressed. Such a feature ensures that the washout rates remain consistently below the Hubble expansion rate, preserving all lepton asymmetry generated in the decay of right-handed neutrinos. We illustrate the feasibility of this mechanism with two example models and show that the observed baryon asymmetry of the Universe can be successfully obtained for right-handed neutrino masses larger than $10^9~{\rm GeV}$ as well as for smaller violation of charge-parity symmetry.
\end{abstract}

\maketitle

\emph{Introduction.}—Leptogenesis is a class of scenarios that provides solutions to the baryon asymmetry of the Universe~(BAU)~\cite{Fukugita:1986hr,Luty:1992un,Davidson:2008bu,Plumacher:1997ru,Buchmuller:2002rq}. In these models, right-handed neutrinos~(RHN), denoted as $N$, are introduced. The RHNs couple to the neutrinos in the Standard Model~(SM) through a Yukawa interaction $\lambda_D\bar\ell_L\tilde H N$, which explains the origin of mass for SM neutrinos~\cite{Davis:1994jw,Super-Kamiokande:1998kpq,KamLAND:2002uet} through the Type-I seesaw mechanism~\cite{Minkowski:1977sc}. A complex $\lambda_D$ matrix coupling of the Yukawa interaction gives rise to the asymmetric, charge-parity~(CP)-violating decay of the RHN, resulting in a lepton asymmetry. Subsequently, this lepton asymmetry is converted to baryon asymmetry through the electroweak sphaleron process. 
 
In the framework of conventional thermal leptogenesis, a major obstacle arises from the strong washout effects that remove the lepton asymmetry. The final BAU is a competition between the asymmetric decay of the RHNs and the washout processes. At temperatures much higher than the thermal leptogenesis scale, the rates of decay and washout processes are smaller than the expansion rate of the Universe. As the temperature decreases, the decay rate of the RHN increases while the washout rates decrease slower than the cosmic expansion rate and eventually surpass it. Then at even lower temperatures, the washout rates are Boltzmann suppressed and drop rapidly below the expansion rate again~\cite{Buchmuller:2004nz,Buchmuller:2002rq}. Typically, the washout effects are strong, and only $\mathcal{O}(10^{-2})$ of the asymmetry generated in RHN decay could survive~\cite{Buchmuller:2004nz,Buchmuller:2002rq}. Assuming a mass hierarchy between three generations of RHNs, the CP-violating phase has an upper bound that is proportional to the mass of the lightest RHN, denoted as $M_1$, known as the Davidson-Ibarra bound~\cite{Davidson:2002qv}. It suggests that, for generating the observed BAU, the  lightest RHN mass is $M_1 \gtrsim 10^{9}$ GeV. Then in the strong washout regime, the typical lightest RHN mass is $M_1 \gtrsim 10^{11}$ GeV.

In this Letter, we propose a mechanism to solve the washout problem by incorporating a dynamical coupling in leptogenesis. In particular, we focus on the strong washout regime of the leptogenesis parameter space. The introduction of temperature-dependent elements is used in different manners to produce dark matter~\cite{Cohen:2008nb, Cui:2011qe, Baker:2016xzo, Baker:2017zwx, Croon:2019ugf, Heurtier:2019beu, Croon:2020ntf} and BAU~\cite{Dutta:2018zkg, Ellis:2019flb, Azatov:2021irb, Baldes:2021vyz, Croon:2022gwq, Huang:2022vkf, Chun:2023ezg, ChoeJo:2023ffp, ChoeJo:2023cnx}. Here, our mechanism exploits the fact that washout processes depend on scattering particle energies in the thermal bath, while the RHN decay process is not affected in the same way. We suppress the coupling $\lambda_D$ at high temperatures to delay the onset of both processes, and restore it to the thermal value when washout processes are kinematically suppressed. Consequently, RHNs decay asymmetrically into the lepton asymmetry while washout rates remain below the Hubble expansion rate, enabling a new leptogenesis scenario. The remainder of this study is arranged as follows: We begin by describing the impact of a temperature-dependent $\lambda_D$ on reaction rates of leptogenesis. We then investigate the enhancement of BAU that can be reached with dynamical couplings. Next, we discuss potential realizations of the mechanism, present our results regarding the RHN mass and CP-violation, and conclude with a final discussion.

\begin{figure*}\centering
\includegraphics[width=\columnwidth]{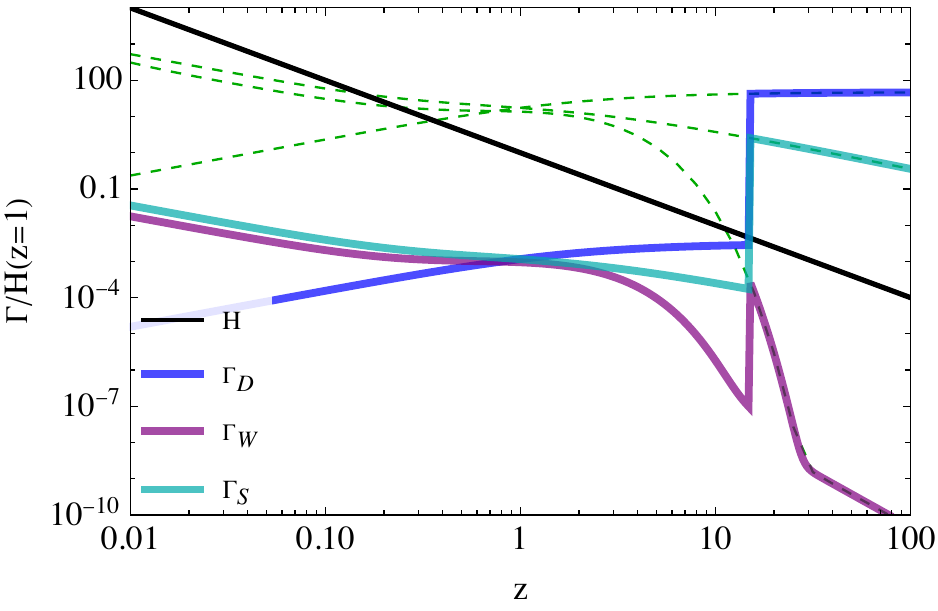}\qquad
\includegraphics[width=\columnwidth]{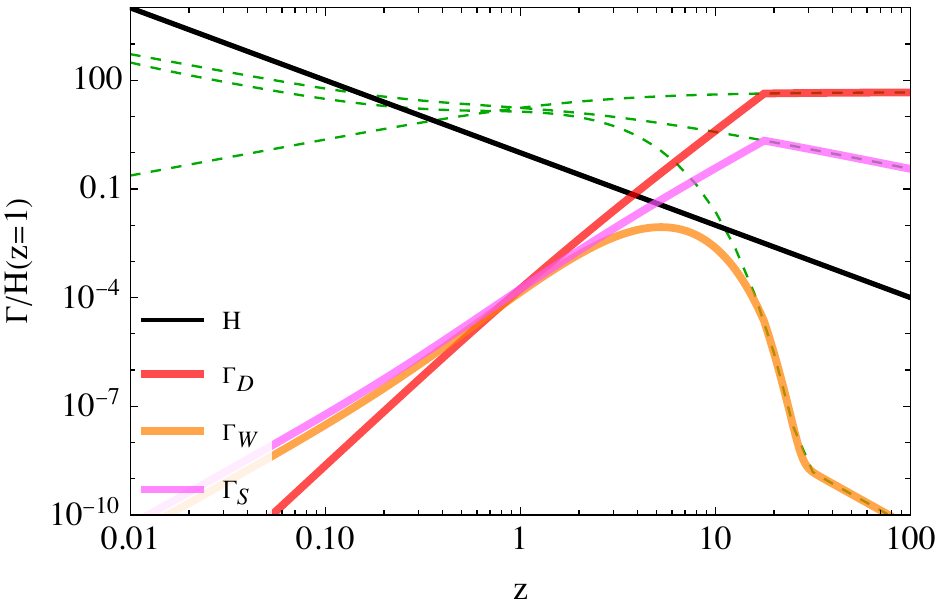}
\caption{Reaction rates of decay, washout, and scattering processes normalized to the Hubble rate at $z=1$. The benchmark neutrino parameters are chosen $M_1=10^{9}~{\rm GeV}$, $\tilde{m}=0.05~{\rm eV}$, and $\bar{m}=0.05~{\rm eV}$. Reaction rates in thermal leptogenesis are shown in dashed green curves. The Hubble rate is shown in the black curve. Left panel: temperature dependence of $\lambda_{D, \rm s}$ follows the step function form, with $z_{\rm s}=15$, $b_{\rm s}\simeq 0.08$. Right panel: temperature dependence of $\lambda_{D, \rm p}$ follows the power law form, with $z_{\rm p}\simeq18$, $a_{\rm p}=2$.}
\label{fig:benchmark}
\end{figure*}

\emph{Temperature-dependent couplings.}—To illustrate the impact of a temperature-dependent $\lambda_D$ on leptogenesis, we introduce two example scenarios, each outlining the variations in $\lambda_D$ as a function of temperature. We assume a mass hierarchy between three generations of RHNs, $M_{2,3}\gg M_1$, and only consider $\lambda_D$ coupling to the lightest RHN particle in this work. Starting with a coupling $\lambda_{D, \rm s}$ that goes through a sudden increase in a step function form as
\bea
\lambda_{D, \rm s}(z)= \begin{dcases} b_{\rm s} \, \lambda_D, & z < z_{\rm s} \\ \,\,\, \lambda_D, & z \geq z_{\rm s} \end{dcases}
\label{eq:steplambda}
\eea
Here $T$ is the temperature of the Universe, and $z \equiv M_1/T$. The constant $b_{\rm s}$ represents the suppression factor at high temperatures before $z_{\rm s}$. At lower temperatures, $\lambda_{D, \rm s}$ aligns with the zero-temperature value $\lambda_D$ for the correct neutrino mass.

The second scenario involves a power law dependence in the coupling $\lambda_{D,\rm p}$ as
\bea
\lambda_{D,\rm p}(z)= \begin{dcases} \lambda_D \, \left( \frac{z}{z_{\rm p}} \right)^{a_{\rm p}}, & z < z_{\rm p} \\ \,\,\, \lambda_D, & z \geq z_{\rm p} \end{dcases}
\label{eq:powerlambda}
\eea
where $a_{\rm p}$ represents the power law index governing the $z$-dependence at high temperature. Similar to Eq.~\eqref{eq:steplambda}, $\lambda_{D,\rm p}$ reverts to $\lambda_D$ at lower temperatures. We note that the proposed mechanism is also viable with other temperature dependence.

The coupling $\lambda_D$ determines reaction rates in leptogenesis. Here we discuss briefly the most relevant processes of RHN decay, scattering, and washout. The decay process $N\to H \ell, \bar{H}\bar{\ell}$ is determined by the intrinsic decay width of the lightest RHN. While the decay violates lepton number by one unit $\Delta L=1$ for generating the lepton asymmetry, the inverse decay process will also contribute to the washout of the existing asymmetry. Another $\Delta L=1$ process is the Higgs-mediated
scattering $N \ell(\bar{\ell})\to\bar{t}q(t\bar{q})$ and $N t(\bar{t})\to\bar{\ell}q(\ell\bar{q})$. The scattering process changes both the number of RHN and the asymmetry. In the end, there are $\Delta L=2$ processes mediated by the RHN,  $\ell\ell\to\bar{H}\bar{H}$, $\bar{\ell}\bar{\ell}\to H H$, and $\ell H \to\bar{\ell}\bar{H}$, which contribute to the washout effect. 

The Boltzmann Equations of the number of lightest RHN $N_1$ and the $B-L$ asymmetry $N_{B-L}$ are as follows, \footnote{In this study, the contribution from CP-violating $\Delta L=1$ scattering processes in generating $N_{B-L}$ is not included, since we aim to resolve the washout issue in the strong washout regime and therefore assume the RHNs are initially in equilibrium. In the most general leptogenesis scenario, a lepton asymmetry can be generated during the production of RHN abundance via Higgs-mediated scattering processes; see, for example,~\cite{Abada:2006ea}. It is noteworthy that the CP-violating scattering can play a role in the weak washout regime and when flavor effects are considered. However, as previously mentioned, their impact is reduced in the strong washout regime due to rapid washout processes. Additionally, in our mechanism, scattering processes that produce RHNs with the normal seesaw Dirac Yukawa coupling are slow at $z \lesssim 1$, and are Boltzmann suppressed when $T \ll M_1$. This indicates that in the dynamical coupling mechanism, the primary source of lepton asymmetry remains the RHN decay process at low temperatures.}
\bea
\frac{dN_1}{dz} &=& -(D+S) \, ( N_1 - N_1^{\rm eq} ), \nonumber\\
\frac{dN_{B-L}}{dz} &=& -\epsilon \, D \, ( N_1 - N_1^{\rm eq} ) - W \, N_{B-L}.
\label{eq:BE}
\eea
The quantities are defined in a comoving volume containing one photon. $N_1^{\rm eq}$ is the equilibrium value of the RHN number. $\epsilon$ is the CP asymmetry in the RHN decay. The rates of decay, scattering, and washout processes are defined as 
\bea
D \equiv \frac{\Gamma_D}{H \, z}, \quad
S \equiv \frac{\Gamma_S}{H \, z}, \quad
W \equiv \frac{\Gamma_W}{H \, z}.
\label{eq:Rate}
\eea
Here $\Gamma_D$ is the RHN decay rate for the generation of asymmetry as well as driving the RHN number to its equilibrium value. $\Gamma_S$ includes rates of $\Delta L=1$ scattering processes, which also contributes to the equilibrium of RHN number. The washout rate $\Gamma_W$ contains the inverse decay of RHN, and the scattering with $\Delta L=1$ and $\Delta L=2$. See the appendix for detailed reaction rates in the conventional leptogenesis scenario. To implement the temperature-dependence in the numerical simulations, we replace $\lambda_D$ with dynamical couplings introduced in Eq.~\eqref{eq:steplambda} and Eq.~\eqref{eq:powerlambda}, and use them to calculate Eq.~\eqref{eq:Rate}. The novel dynamics will not only rescale reaction rates with their proportionality in $\lambda_D$, but also set back the time when interactions come into equilibrium, thus leading to an additional kinematical suppression in the washout effect.

\begin{figure}
\centering
\includegraphics[width=\columnwidth]{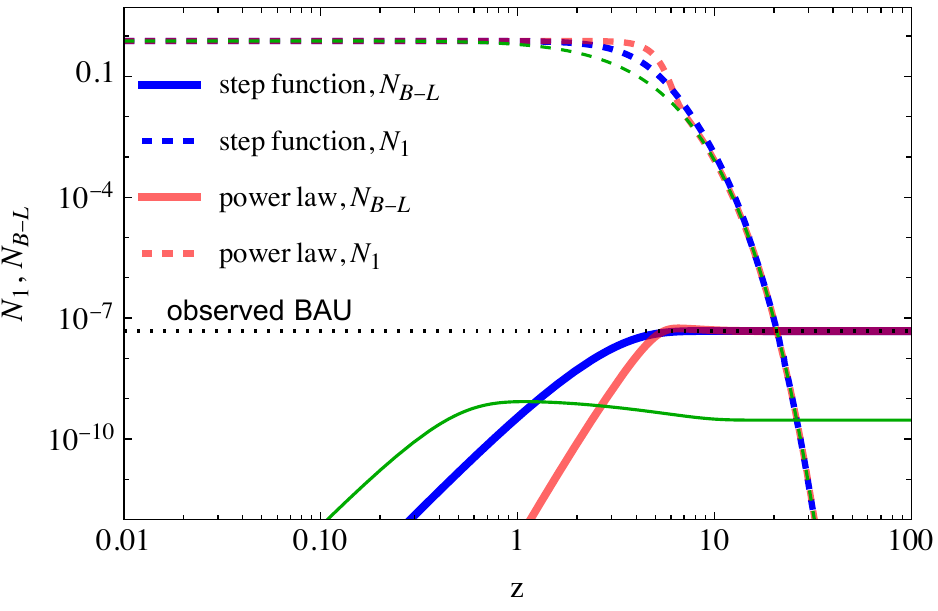}
\caption{Evolution of RHN number $N_1$ (dashed) and $B-L$ number $N_{B-L}$ (solid) in the comoving volume. The benchmark parameters are the same as in Fig.~\ref{fig:benchmark}. The blue and red curves represent the step function case and the power law case, while the thermal leptogenesis case is shown in green curves for comparison. The $N_{B-L}$ needed to explain the observed BAU is shown in the black dotted line.}
\label{fig:benchmarkNumber}
\end{figure}

\begin{figure*}
\centering
\includegraphics[width=\columnwidth]{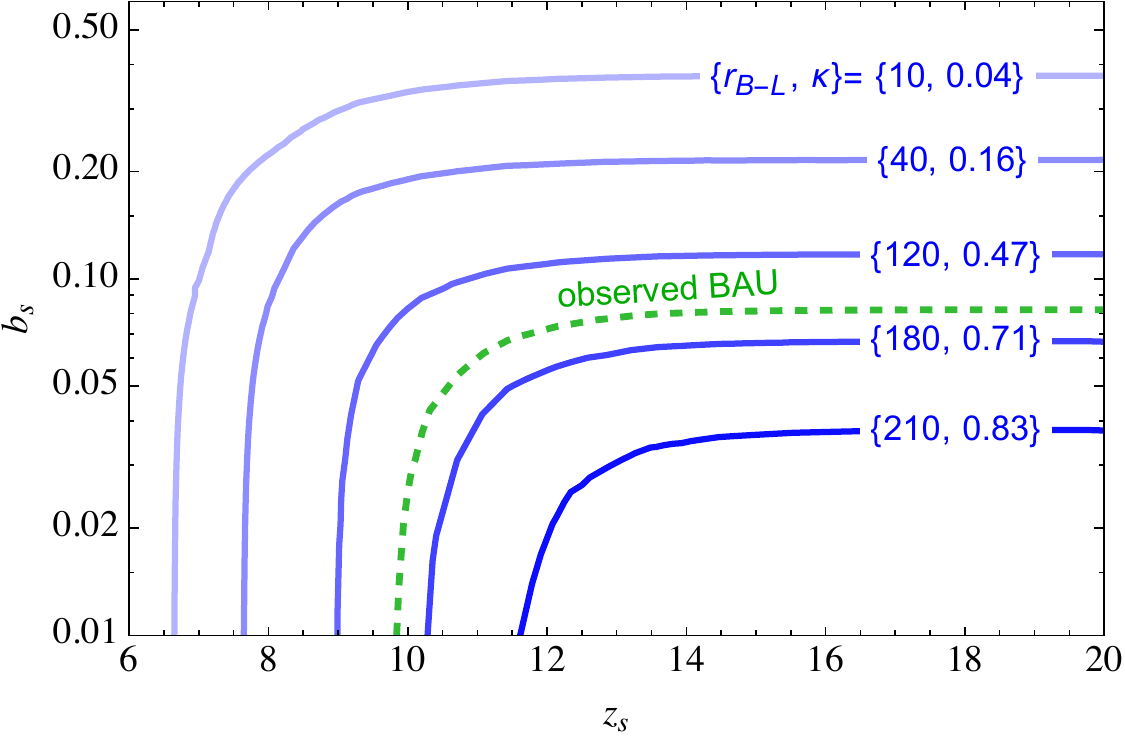}\qquad
\includegraphics[width=\columnwidth]{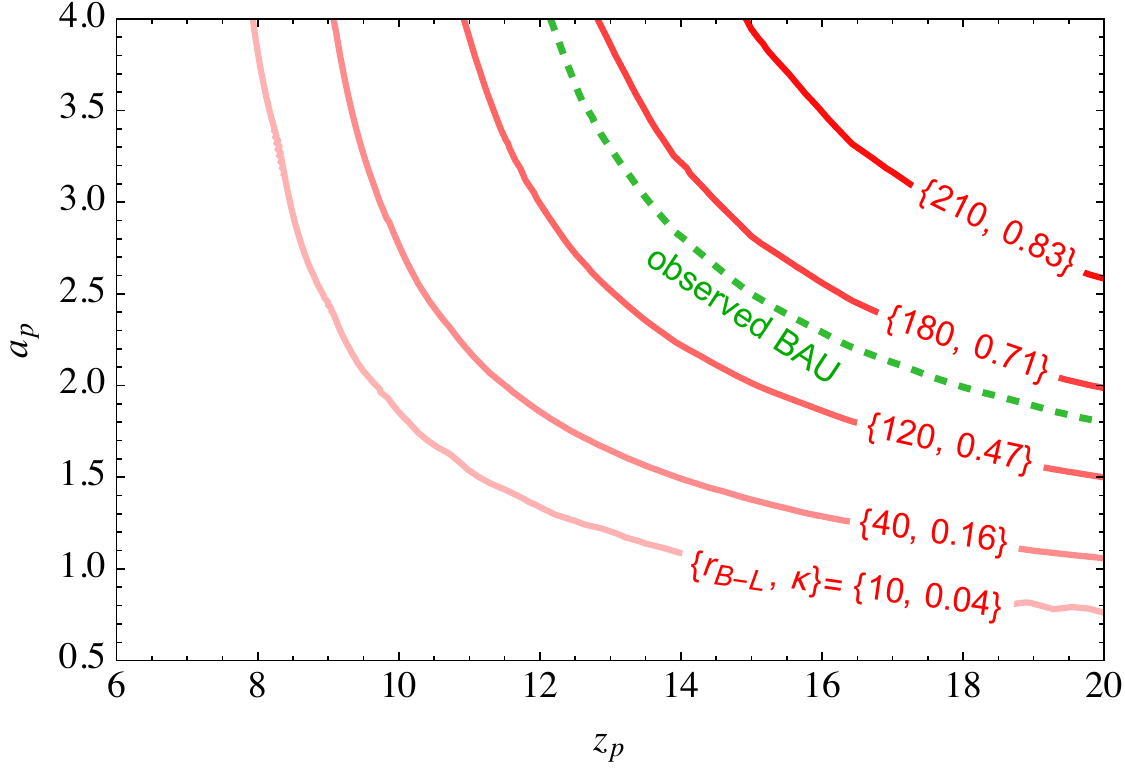}
\caption{Enhancement ratio $r_{B-L}$ and the efficiency factor $\kappa$ obtained with different temperature-dependence parameters, $\{z_{\rm s}, b_{\rm s}\}$ in the step function scenario (left panel) and $\{z_{\rm p}, a_{\rm p}\}$ in the power law scenario (right panel). The RHN parameters are kept to match benchmark models in previous figures. The blue and red lines show model parameters that lead to constant enhancement values labeled in text for the step function scenario and power law scenario respectively. The $r_{B-L}$ that generates the observed BAU is shown with green dashed lines in both panels.}
\label{fig:switchparam}
\end{figure*}

In Fig.~\ref{fig:benchmark}, we illustrate the normalized reaction rates of the RHN decay, the washout of lepton asymmetry, and the RHN scattering with SM particles in the thermal bath within different leptogenesis scenarios. Rates of thermal leptogenesis are shown in dashed green curves, while rates with dynamical couplings are in solid color curves. Additionally, we depict the normalized Hubble rate in black to show active epochs of processes whose rates are above the Hubble rate. The benchmark parameters are chosen as follows: The RHN mass $M_1=10^{9}~{\rm GeV}$. The effective lightest neutrino mass $\tilde{m}$ is defined as $\lambda_D^2 v_h^2/M_1$, where $v_h =174$~GeV is the Higgs vacuum expectation value~(vev). We take $\tilde{m}=0.05~{\rm eV}$ to be much larger than the equilibrium neutrino mass $m_\star\simeq 1.1\times10^{-3}~{\rm eV}$, such that the dynamical coupling mechanism is validated in the strong washout regime $\tilde{m}\gg m_\star$~\cite{Buchmuller:2003gz}. $\bar{m}$ is defined as $\bar{m}^2 = tr(m_{\nu}^{\dagger}m_{\nu})$, and we take $\bar{m}=0.05~{\rm eV}$. We also assume that RHNs are initially in thermal equilibrium. This equilibrium state can be attained by coupling RHNs to particles that are thermalized with the SM from the beginning.

The incorporation of temperature-dependent couplings can significantly modify the leptogenesis paradigm. Fig.~\ref{fig:benchmark} show effects of dynamical couplings in Eq.~\eqref{eq:steplambda} (left) and in Eq.~\eqref{eq:powerlambda} (right) respectively. The step-function coupling is suppressed by a factor of $0.08$ for $z<15$. Similarly, the power-law coupling has a $T^{-2}$ suppression for $z<18$. As the couplings are practically turned off at high temperatures, all three reaction rates consistently remain lower than the Hubble expansion rate when $z \lesssim 1$, in contrast to the thermal leptogenesis case where processes become efficient after $z\simeq 0.2$. As the temperature decreases further, the RHN decay rate increases to exceed the Hubble rate when the coupling reverts to $\lambda_D$, such that the lepton asymmetry is generated. However, even with this late-time increase of couplings, washout rates are substantially Boltzmann suppressed as the temperature has dropped well below $M_1$ at this stage. The core feature of our mechanism is this absence of active washout process, which is inherent for conventional thermal leptogenesis, and its impact on the generation of BAU is discussed in the following.

The generated BAU results from the competition between the RHN decay and the washout processes that act in opposition to it. We solve the standard Boltzmann Equations in Eq.~\eqref{eq:BE} for the time evolution of RHN number $N_1$ and the $B-L$ number $N_{B-L}$, and show the results in Fig.~\ref{fig:benchmarkNumber}.
The CP-violation $\epsilon=10^{-7}$ is assumed to saturate the Davison-Ibarra bound. Since the washout rate in thermal leptogenesis is much larger than the Hubble rate between $z\simeq 0.2$ and $z\simeq 10$, the final $N_{B-L}$ (green solid) is about two orders of magnitude lower than the value for generating the observed BAU \cite{Workman:2022ynf}, as indicated by the black dotted line. Once dynamical coupling suppressions are introduced to reduce washout effects, the resultant $N_{B-L}$ from the step-function dependence (blue solid) and the power-law dependence (red solid) are about two orders of magnitude greater than that in the thermal leptogenesis benchmark point, eventually aligning with the observed value. 

In Fig.~\ref{fig:switchparam}, we present enhancement in $N_{B-L}$, defined as $r_{B-L}\equiv N_{B-L}/N_{B-L}^{\rm thermal}$, achieved with temperature-dependent couplings for the same benchmark parameters, assuming RHNs are initially in equilibrium. For the enhancement in $r_{B-L}$, we also show the corresponding efficiency factor of the final asymmetry $\kappa$ introduced in~\cite{Barbieri:1999ma}. The efficiency factor $\kappa\simeq1$ in the weak washout regime, and is suppressed in parameter space where washout becomes relevant. The green dashed line represents the enhancement needed for the observed BAU. In the left panel, we show the step function scenario. Greater suppression in the coupling due to smaller $b_{\rm s}$ values leads to a more substantial reduction of washout effects, which in turn results in a larger enhancement over thermal leptogenesis. The maximum enhancement reaches about 200 and the maximum efficiency factor is close to $1$, corresponding to the case of highly suppressed washout. As the suppression persists to higher $z_{\rm s}$ values, signifying a delayed end to the dynamical effect, the subsequent Boltzmann suppression in washout effectively retains more lepton asymmetry at lower temperatures. The right panel displays a similar trend in the power law scenario. Higher power law indices $a_{\rm p}$ are associated with larger suppression of washout rates, and greater $z_{\rm p}$ values extend the duration of the suppression phase. Both factors contribute to a more pronounced enhancements of BAU. 

\emph{Possible realizations and probes.}—We explore possible realizations of temperature-dependent couplings. The novel dynamics proposed can be achieved by introducing a new scalar field $S$ that interacts as 
\begin{equation}
    \mathcal{L
    }\supset -\left(\operatorname*{\sum_{\textit{i,\,j}}}\lambda^{i,j}_{D,0}\bar{l}_L^i\tilde{H}N^{j} + \operatorname*{\sum_{\textit{i,\,j}}}\tilde{\lambda}_D^{i,j}\frac{S}{\Lambda_s}\bar{l}_L^i\tilde{H}N^{j}\right)+h.c.
    \label{eq:newscalar}
\end{equation}
Here we also keep the normal Dirac Yukawa term. If the scalar field $S$ acquires a vev $v_s$, its contribution to the second term in Eq.~\eqref{eq:newscalar} becomes $(\tilde{\lambda}_D^{i,j} v_s/\Lambda_s) \, \bar{l}_L^i\tilde{H}N^{j}$ and enables an effective Dirac Yukawa coupling in the form $\lambda_{D,0}+\tilde{\lambda}_D v_s/\Lambda_s$, which evolves with $v_s$. Such effective couplings are dynamical during the expansion of the Universe since the value of $v_s$ depends on the temperature. To study the impact of dynamical couplings, we replace the coupling $\lambda_D$ of conventional thermal leptogenesis with effective couplings obtained with Eq.~\eqref{eq:newscalar} to account for the $S$ contributions in the decay and scattering of RHNs. Note that the value of the dynamical coupling is $\lambda_{D,0}$ before $S$ obtaining a considerable vev, and in our mechanism $\lambda_{D,0}$ is smaller than that in the conventional leptogenesis scenario in order to suppress washout processes. Observe that, at $z\gtrsim1$, the dynamical coupling only restores to the conventional leptogenesis value, thus it does not necessarily lead to higher interaction rates compared to thermal leptogenesis. The crucial role of the temperature dependence is to suppress the washout rate and enable the efficiency factor $\kappa$ to reach order unity.

The new scalar $S$ can introduce additional interactions for the RHN. However, these interactions are model-dependent, and we outline possible cases below. Considering the singe flavor regime as an example, the $(\tilde{\lambda}_D/\Lambda_s)\, S\, \bar{l}_L \tilde{H} N$ interaction can keep RHNs in equilibrium with the SM sector via scattering with $S$, leptons, and the Higgs in the thermal bath. At a given time $z$, the equilibrium condition for $N$ is obtained by requiring the reaction rate to be larger than the Hubble rate, which translates to $\tilde{\lambda}_D\gtrsim 0.15 \, (z/0.1)^{\frac{1}{2}}\,(10^{9}~{\rm GeV}/M_1)^{\frac{1}{2}}\,(\Lambda_s/10^{13}~{\rm GeV})$ for the choice of $\tilde{\lambda}_D$ parameter. Although the reaction rate from $(\tilde{\lambda}_D/\Lambda_s)\,S\,\bar{l}_L \tilde{H} N$ becomes Boltzmann-suppressed once the temperature falls below the mass of $S$, keep in mind that this merely leads to separated evolutions of the RHN sector and the SM sector. Before $N$ decays, its comoving number is only significantly altered in the presence of a model-dependent, large entropy injection into the SM sector. Otherwise, the equilibrium condition can be reliably used for the initial $N$ abundance when solving the Boltzmann equations in Eq.~\eqref{eq:BE} after $S$ decouples. Another effect of $S$ is that it can mediate RHN scattering with SM particles via loop-level processes. We found that the $S$-loop mediated processes are decoupled as long as $\tilde{\lambda}_D\lesssim 2.8\times10^{2}\,z^{\frac{3}{4}}\,(10^9~{\rm GeV}/M_1)^{\frac{3}{4}}\,(\Lambda_s/10^{13}~{\rm GeV})$. This suggests that $S$ influences leptogenesis primarily through its vev in the dynamical coupling and its scattering, which keeps the RHN in equilibrium. However, its effect on the RHN abundance is minor for $z \gtrsim 1$, except for potential enhancements in specifically designed model realizations. More importantly, the deviation of the initial $N_1$ from thermal equilibrium does not affect the $\kappa \simeq 1$ achieved by suppressing the washout rate. Therefore, we do not introduce new interactions to the Boltzmann equations other than assuming that $N$ is initially in equilibrium, but focus on the suppression of the washout rate in our numerical analysis.

\begin{figure*}
\centering
\includegraphics[width=\columnwidth]{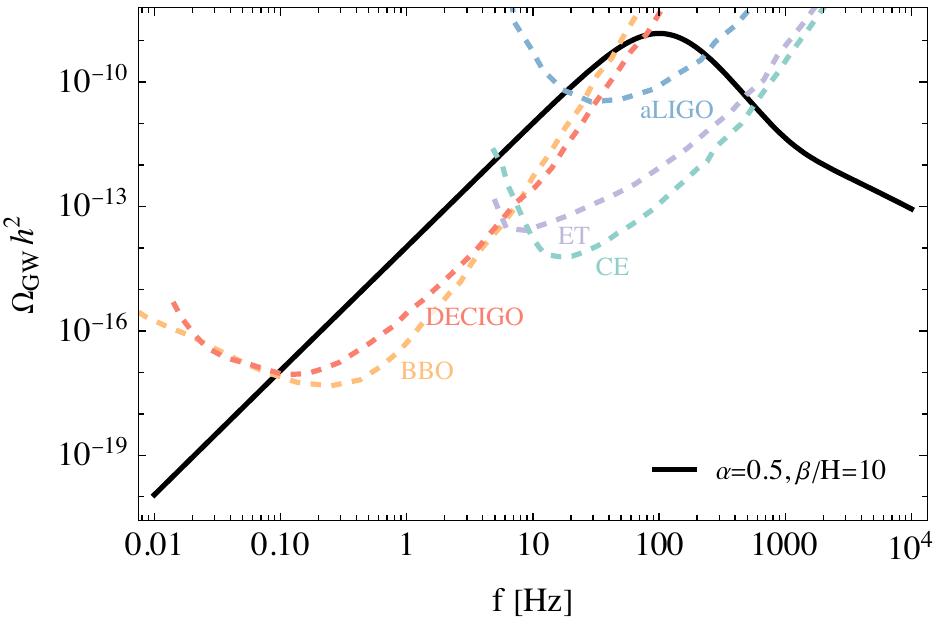}\qquad
\includegraphics[width=\columnwidth]{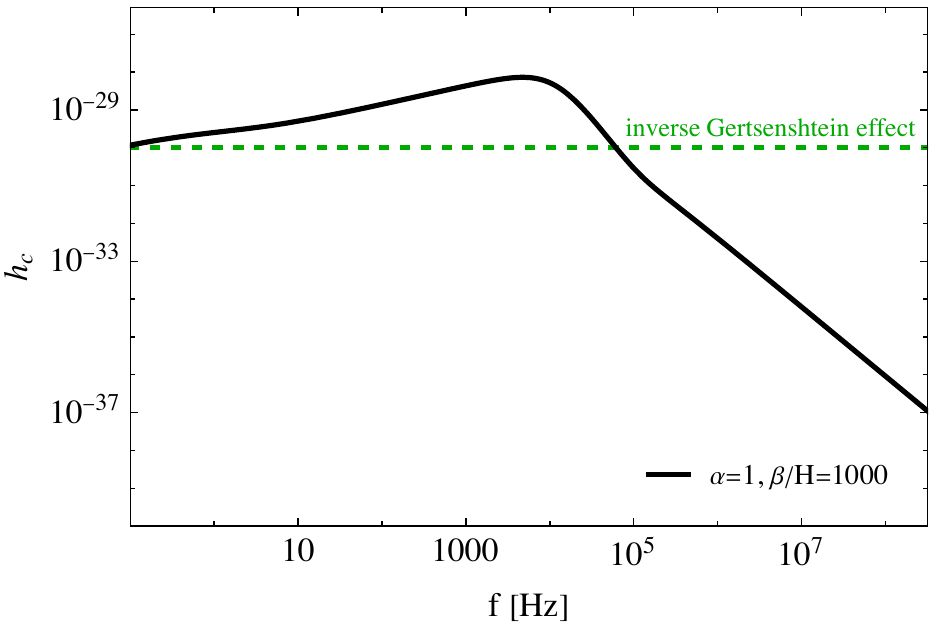}
\caption{GW signal from a first order PT model realization of the mechanism. We choose a PT temperature $T_p = 6.7\times 10^7~{\rm GeV}$ corresponds to $z_s=15$ for the RHN mass $M_1=10^{9}~{\rm GeV}$. The PT bubble wall velocity is $v_w=0.6$. Left panel: the black curve shows the GW spectrum corresponding to $\alpha = 0.5$, and $\beta/H = 10$. Color curves show that future GW observations can probe the PT model via GWs. Right panel: the black curve shows the characteristic strain for $\alpha = 1$, and $\beta/H = 1000$. The high-frequency GW signal is expected to be probed with the inverse Gertsenshtein effect, whose anticipated sensitivity is indicated with the green dashed line.}
\label{fig:GWBM}
\end{figure*}

A scalar field with temperature-dependent vev can be realized in many models. In the previous discussion, we discussed two scenarios of the effective Dirac Yukawa coupling being a step function and a power law function. The step function case can be easily realized in a first-order phase transition. When the temperature is above the phase transition~(PT) temperature $T_p$, which corresponds to $M_1/z_{\rm s}$ in Eq.~\eqref{eq:steplambda}, the vev of $S$ is zero. The Dirac Yukawa $\lambda_{D,0}$ is the temperature-independent contribution corresponding to $b_{\rm s} \lambda_D$. RHNs remain in equilibrium at high temperatures due to their thermal contact with $S$. After the PT, the scalar field $S$ acquires a vev $v_s$, and the effective Yukawa coupling becomes $\lambda_{D,0} + \tilde{\lambda}_D v_s/\Lambda_s$, whose value equals to $\lambda_D$. Our benchmarks show that the Boltzmann suppression on washout rates begins to take effect when $z_{\rm s} \gsim 10$, which corresponds to PT temperatures $T_p \lsim 0.1  M_1$, as observable in the left panel of Fig.~\ref{fig:switchparam} where contours of constant $r_{B-L}$ tend to flatten in regions of large $z_{\rm s}$.

A strong first-order PT is easily realized for a dark sector particle. For example, the most general renormalizaible effective potential of the singlet takes the form of
\begin{equation}
    V(S,T) = A(T^2-T_0^2)S^2 - (BT+C)S^3+D S^4,
\end{equation}
in which $A,\, T_0,\, B, \,C$ and $D$ are free parameters. The temperature dependent piece can be generated from its coupling to other particles. The mixing between the singlet $S$ and the SM Higgs can be allowed. However, due to the large mass hierarchy between the singlet and Higgs mass in this scenario, the singlet field has little impact on leptogenesis other than providing a dynamical coupling.

Model building for the power law scenario is inspiringly more diverse. One simple example is to couple the RHN to a new field $S$ whose potential is similar to the relaxion potential~\cite{Graham:2015cka}. At high temperatures, the field value of $S$ increases as it rolls down in time, until the process is stopped by a potential barrier. The temperature where the vev of $S$ freezes corresponds to $M_1/z_{\rm p}$ in Eq.~\eqref{eq:powerlambda}. The detailed temperature dependence in $\lambda_{D,\rm p}$ scanned by $S$ is determined by the rolling dynamics along the scalar potential. In the end, the value of $\lambda_{D,\rm p}$ at $z_{\rm p}$, which determines both the BAU and the SM neutrino mass, is dynamically selected to match the thermal coupling at low temperatures. Without going into details of constructing the scalar potential model, our benchmark analysis in the right panel of Fig.~\ref{fig:switchparam} reveals that an overall power index of $a_{\rm p} \gtrsim 1.8$ is needed for $z_{\rm p}\leq 20$.

Here we use a first order PT model as an illustrative example to demonstrate the probe of the mechanism. In particular, we selected $z_s = 15$ for $M_1=10^9~{\rm GeV}$, which implies a PT occuring at temperature $T_p \simeq 6.7 \times 10^7~{\rm GeV}$. A first order PT is expected to generate a stochastic background of gravitational waves~(GWs). The GW spectrum is determined by two key parameters of the PT model. The first parameter, $\alpha$ represents the ratio between the latent heat released during the PT and the radiation energy density at the temperature at which the PT occurs. The second one, commonly referred to as $\beta$, sets the inverse timescale associated with the PT duration. Both parameters are determined upon specification of the potential of $S$. In Fig.~\ref{fig:GWBM}, we chose $\alpha = 0.5$, $\beta/H = 10$, in which $H$ is the Hubble parameter at the time of the PT, and plot corresponding GW spectrum. We also show the expected sensitivity curves for the space-based laser Interferometers DECIGO~\cite{Decigo}, and BBO~\cite{BBO:2005nr}, and the ground-based interferometers aLIGO~\cite{aLIGO:2019vic}, ET~\cite{ET:2010id,ET:2012jk}, and CE~\cite{2019BAAS...51g..35R} in the left panel of Fig.~\ref{fig:GWBM}. For this benchmark, the peak frequency is about 100 Hz, and corresponding signal can be probe by aLIGO, ET, and CE. Notably, $\beta/H = 10$ suggests a relatively slow PT, resulting in less sharp changes in the temperature-dependent coupling compared to a step function. However, the washout effects remain Boltzmann suppressed as long as $T_p$ is much lower than $M_1$. The decay and scattering rates can be determined with the time scale of a slow PT included in the simulation, and the effect on the final $B-L$ yield can be compensated by a more suppressed $b_s$. Subsequently, we explore another benchmark with $\alpha =1$, and $\beta/H =1000$. In this case, a larger $\beta/H$ shifts the peak GW frequency to higher values. We plot the strain for this benchmark point in the right panel of Fig.~\ref{fig:GWBM}, where the shaded region represents the anticipated sensitivity of enhanced magnetic conversion for high-frequency GW searches using the inverse Gertsenshtein effect~\cite{Franciolini:2022htd,Ringwald:2020ist}. Future high-frequency GW searches will also probe PTs with higher $T_p$ values that realize dynamical couplings for larger $M_1$ masses. These two benchmarks illustrate that the PT model proposed in this scenario for generating the correct BAU can be detected in future experiments.

\emph{Results.}—The temperature dependence in the $\lambda_{D}$ coupling significantly reduces the dilution of $N_{B-L}$ due to washout, opening up new parameter space for realizing leptogenesis. Notably, this allows for a lighter minimum RHN mass compared to the previous limitations set by the Davidson-Ibarra bound within the strong washout regime, while still being able to generate an adequate BAU. Furthermore, the lepton asymmetry arising from RHN decay is directly proportional to the CP-violation parameter $\epsilon$, and is therefore contingent upon specific model configurations to realize large $\epsilon$ values. When the washout effect is reduced, the necessary value of $\epsilon$ correspondingly decreases in proportion to enable more model space.

\begin{figure}\centering
\includegraphics[width=\columnwidth]{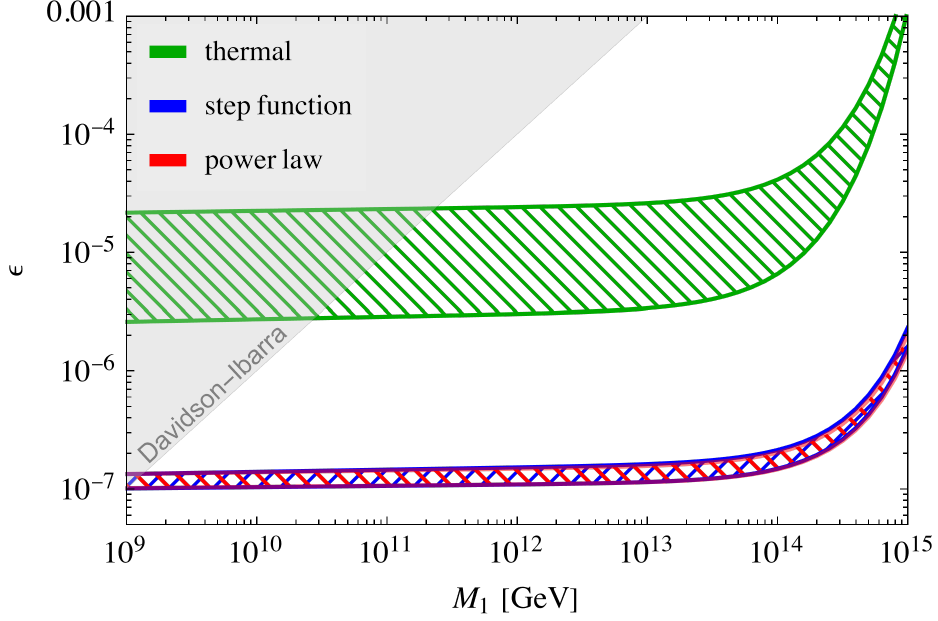}
\caption{CP-violation parameter $\epsilon$ for generating the observed BAU with different RHN mass $M_1$. The green, blue, and red bands show ranges of $\epsilon$ needed for $8~{\rm meV} \leq \tilde{m} \leq 50 ~{\rm meV}$ in thermal leptogenesis, step function, and power law scenarios. Other parameters are consistent with Fig.~\ref{fig:benchmark}. The region constrained by the Davidson-Ibarra bound is shown in the gray-shaded area. }
\label{fig:CPV}
\end{figure}

In Fig.~\ref{fig:CPV}, we present the derived CP-violation criteria required for successful leptogenesis within an extensive lightest RHN mass range. Without loss of generality, we maintain the benchmark from Fig.~\ref{fig:benchmark} for consistency, except for permitting the effective neutrino mass to range from $8~{\rm meV} \leq \tilde{m} \leq 50 ~{\rm meV}$ to showcase results with varying washout strength in the strong washout limit. The range of $\tilde{m}$ is motivated by neutrino oscillation measurements with solar neutrinos and atmospheric neutrinos~\cite{Workman:2022ynf}. The step function scenario is shown in the blue-hatched band, and the power law scenario is in the red-hatched band, while the comparative thermal case is represented in green. Our dynamical mechanism significantly relaxes the previous demands on CP-violation in the strong washout regime, with a reduction in $\epsilon$ by about two orders of magnitude for the largest chosen $\tilde{m}$ value (upper edge of the green-hatched band) within the range $M_1 \lesssim 10^{14}~{\rm GeV}$. This leads to a minimum mass for the lightest RHN of approximately $10^{9}~{\rm GeV}$, which is $\mathcal{O}(100)$ times lower than the thermal case for the given range of $\tilde{m}$. At heavier masses, the RHN-mediated off-shell washout process becomes increasingly effective in removing the generated asymmetry, causing all color bands to extend into larger $\epsilon$ regions. Observe that the difference from the thermal case remains clearly discernible despite the shift, underscoring the impacts of temperature-dependent couplings.

Hatched areas in Fig.~\ref{fig:CPV} visually represents the general outcome of dynamically suppressed couplings with their shapes. The parameter spaces of two temperature-dependence models largely overlap, because the final $N_{B-L}$ is primarily determined by the RHN number $N_1$, as long as $\lambda_D$ is sufficiently suppressed at high temperatures. This is equivalent to achieving an efficiency factor $\kappa$ that saturates at unity. Thus, comparable levels of CP-violation are required to generate the BAU. It also explains the narrow widths of the blue and red bands, as the increase in $\epsilon$ demonstrates only a mild dependence on $\tilde{m}$, contrary to thermal leptogenesis, where $\epsilon$ increases by almost an order of magnitude for the indicated rise in $\tilde{m}$. Therefore, the proposed mechanism is viable for various suppression models beyond current examples.

We comment on flavor effects in thermal leptogenesis. Our analysis is performed in the unflavored approximation where coherence amongst lepton flavors are preserved for RHN decay final states. In a more generalized parameter space, charged-lepton flavor effects become relevant when BAU is generated at low temperatures. This is because interactions between charged leptons and Higgs enter equilibrium as the Universe expands and could break the coherence in the flavor space before washout takes place, thus leading to BAU generation in a flavored leptogenesis scenario~\cite{Abada:2006fw, Nardi:2006fx, Abada:2006ea, Blanchet:2006be, Pascoli:2006ie, DeSimone:2006nrs}. The lepton flavor effect is recently reviewed in~\cite{Dev:2017trv}. Here we briefly summarize the impact on the case of hierarchical RHN masses assumed in this study. There are three relevant flavor effect regimes determined by the typical leptogenesis scale $M_1$. In the high scale limit $M_1\gtrsim 5\times 10^{11}~{\rm GeV}$, the flavor effect is negligible since lepton interactions are out of equilibrium. The two-flavor regime occurs for $5\times10^{8}~{\rm GeV} \lesssim M_1\lesssim 5\times10^{11}~{\rm GeV}$ where asymmetries in the $\tau$ flavor and $e+\mu$ flavor can be tracked separately. This two-flavor regime may deviate significantly from the unflavored approximation, potentially lowering the minimum mass of the lightest RHN, though some parameter tuning might be required. Previous studies found that the bound on $M_1$ can be relaxed to about $10^8~{\rm GeV}$ given a mild tuning in the seesaw model and lowered $M_2$ values~\cite{Blanchet:2008pw}. A more substantial effect arises when the CP asymmetry needed for generating BAU stems completely from the neutrino mixing matrix. This predicts that the leptogenesis CP-violating parameters are connected to the Dirac or Majorana phases in the lepton sector, potentially observable at low energies neutrino oscillation experiments~\cite{Pascoli:2006ci, Anisimov:2007mw, Molinaro:2007uv, Molinaro:2008cw, Molinaro:2009lud}. For example, the connection of the lightest neutrino mass, $m_1$ in the normal mass ordering and $m_3$ in the inverted mass ordering, to flavor effects is studied in~\cite{Molinaro:2007uv} using a seesaw model setup similar to this study. Notably, the dynamical coupling in our work is primarily embedded in high-scale models, while the flavor contributions can be introduced at low energy scales, meaning that both mechanisms can in principle be implemented simultaneously within a consistent model. The last case, $M_1 \lesssim 5\times10^{8}~{\rm GeV}$, is the three-flavor regime where both the $\tau$ interaction and the $\mu$ interaction are efficient. Existing studies found that the BAU in the three-flavor regime is small, unless a quasi-degenerate RHN spectrum or a fine-tuned parameter region is introduced. Finally, we direct readers to~\cite{Dev:2017trv} for discussions on flavor effects in leptogenesis beyond the hierarchical RHN mass assumption.

\emph{Discussion.}—In this Letter, we introduced temperature dependence into the RHN coupling $\lambda_D$, as a new approach to effectively reduce washout effects in leptogenesis. In this scenario, $\lambda_D$ experiences significant suppression at high temperatures, resulting in the simultaneous suppression of both RHN decays and washout processes. As the temperature decreases to below the RHN mass, $\lambda_D$ restores to the thermal leptogenesis value, generating the lepton asymmetry through RHN decays while washout processes are already Boltzmann suppressed. Through this temperature-dependent coupling, the washout effects are highly suppressed, ensuring the survival of all generated asymmetry, hence accounting for the observed BAU.

While this study aims to provide a first attempt to resolve the washout issue with dynamical couplings in the strong washout regime of leptogenesis, there are further model building aspects that can be incorporated in the future. We briefly comment on these possibilities. Firstly, the RHN flavor effect and charged lepton flavor effect are not included in this study. While the RHN flavor effect is minor for the assumed hierarchical RHN mass relation, lepton flavor effects could potentially enhance BAU generation, especially for the lightest RHN mass in the two-flavor region previously discussed. Secondly, although the CP-violation parameter is chosen to be a constant in this model, one can in principle choose the complex neutrino coupling matrix $\lambda_D$ such that the $\epsilon$ parameter also evolves dynamically. A possible scenario is that when the temporary enhancement in $\epsilon$ is introduced at high temperatures, sufficient $B-L$ asymmetry can be generated, even if the RHN abundance is small or the washout remains active. Although the detailed model building of dynamical CP violation is beyond the scope of this work, we note that a consistent model should satisfy constraints on contributions to leptonic CP-violation phases at low energies~\cite{Joshipura:2001ui, Pascoli:2006ie, Pascoli:2006ci, Moffat:2018smo}. Lastly, the dynamical evolution during the generation of BAU can be achieved within neutrino mass models. For example, a seesaw model with dynamical scales is studied in~\cite{AristizabalSierra:2014uzi} where the BAU is enhanced in the weak washout regime without being affected by the initial RHN abundance.

In summary, we introduced a new mechanism that can significantly open up parameter space for leptogenesis. For hierarchical RHN masses, the observed BAU is produced for $M_1 \gsim 10^9~{\rm GeV}$. In particular, our mechanism allows the parameter space $10^9~{\rm GeV} \lsim M_1 \lsim 10^{11}~{\rm GeV}$ in which conventional leptogenesis suffers from strong washout effects. For heavier RHNs, the new method requires smaller CP-violation compared to the Davison-Ibara bound, thereby allowing more versatility in constructing RHN models that link heavy mass scales to phenomenological searches for CP-violation at lower energy scales. Future work dedicated to developing detailed temperature-dependent coupling models holds promising prospects. Additionally, exploring the cosmological evolution within dynamical coupling models is expected to yield complementary signals, such as the aforementioned GWs from PT models, for probing this mechanism. Our findings highlight that dynamical coupling introduces a fresh and compelling pathway in the landscape of leptogenesis.

\medskip
\begin{acknowledgments}
PH is supported by the National Science Foundation under grant number PHY-2112680. TX is supported by DOE grant DE-SC0009956. We thank the Mitchell Institute for Fundamental Physics and Astronomy for providing the platform where initial discussions for this work were developed. We thank the Center for Theoretical Underground Physics and Related Areas (CETUP*), The Institute for Underground Science at Sanford Underground Research Facility (SURF), and the South Dakota Science and Technology Authority for hospitality and financial support, as well as for providing a stimulating environment while this work was in progress.
\end{acknowledgments}

\appendix
\section{Leptogenesis Reaction Rates}
\label{sec:ReactionRates}
In this appendix, we summary reaction rates for leptogenesis following~\cite{Buchmuller:2004nz}, and discuss how to modify the rates to include the dynamical coupling effects. We calculate the leptogenesis process in the standard Friedmann-Robertson-Walker cosmology. The Hubble expansion rate is determined by the energy density
\bea
H^2 = \frac{8 \pi G}{3} \rho.
\eea
The comoving number densities of the lightest RHN $N_1(z)$ and the $B-L$ asymmetry $N_{B-L}(z)$, as a function of $z\equiv M_1/T$, are calculated with the set of differential equations in Eq.~\eqref{eq:BE}. The reaction rates in the equations for decay $D=\Gamma_D/H z$, washout $W=\Gamma_W/H z$, and scattering $S=\Gamma_S/H z$ are defined as follows. 

The decay width of the lightest RHN is 
\bea
\Gamma_D=\frac{\lambda_D^2}{8\pi} \, M_1 \, \frac{K_1(z)}{K_2(z)}.
\eea
Here $K_1$ and $K_2$ are the Bessel functions of the first and the second kind. The coupling $\lambda_D$ is related to the effective lightest neutrino mass $\tilde{m}$, the lightest RHN mass $M_1$, and the Higgs vev $v_h$ as
\bea
\lambda_D^2=\frac{\tilde{m}\, M_1}{v_{h}^2}.
\eea

The total washout rate includes contributions from different processes. We can write down the washout rate as 
\bea
W=W_0+\Delta W,
\eea
where $W_0$ only depends on $\tilde{m}$, and $\Delta W$ depends on $M_1$. The first contribution term can be written as 
\bea
W_0=W_{ID}+W_{\Delta L=1}.
\eea
The $W_{ID}$ is from the inverse decay process of the RHN, and its rate is 
\bea
W_{ID}=\frac{1}{4} \, K \, z^3 \, K_1(z).
\eea
The parameter $K$ is defined with the decay rate at temperatures much lower than the leptogenesis scale, indicating whether the RHN can decay in equilibrium, 
\bea
K=\frac{\Gamma_D(z=\infty)}{H(z=1)}.
\eea
For the parameters used in the main text, $K\gg1$ such that the benchmarks are in the strong washout regime. The second term in $W_0$ is from the washout induced by $\Delta L=1$ scattering processes $N \ell(\bar{\ell})\to\bar{t}q(t\bar{q})$ and $N t(\bar{t})\to\bar{\ell}q(\ell\bar{q})$, mediated by the Higgs. The washout rate is directly related to the scattering rate by
\bea
W_{\Delta L=1}=2\,\frac{W_{ID}}{D} \, \left(\frac{N_1}{N_l^{\rm eq}}\, S_{H,s} + 2 S_{H,t}\right).
\eea
Here the number density $N_1$ of the lightest RHN and the equilibrium number density $N_l^{\rm eq}$ of SM leptons are used. Finally, the washout contribution from $\Delta L=2$ processes $\ell\ell\to\bar{H}\bar{H}$, $\bar{\ell}\bar{\ell}\to H H$, and $\ell H \to\bar{\ell}\bar{H}$ with the RHN in the mediator gives the rate $\Delta W$,
\bea
\Delta W \simeq \frac{\omega}{z^2} \, \left( \frac{M_1}{10^{10}~{\rm GeV}} \right) \, \left( \frac{\bar{m}}{{\rm eV}} \right).
\eea
The value of $\omega \simeq 0.186$ is independent of RHN parameters.

In the end, we include the rate for the scattering processes that are responsible for both the thermalisation of RHN and the washout of $B-L$ asymmetry. In this study, we only include the scattering involving quarks. The scattering rate consists of the $t$-channel term $S_{H,t}$ and the $s$-channel term $S_{H,s}$,
\bea
S=2\, S_{H,s} + 4\, S_{H,t}.
\eea
We refer to Appendix B of~\cite{Buchmuller:2004nz} and reference~\cite{Buchmuller:2002rq} for the explicit form of $S_{H,s}$ and $S_{H,t}$ that are calculated with the reduced scattering cross sections.

In the proposed dynamical coupling scenario, the RHN coupling $\lambda_D$ is promoted from a constant to a function of $z$, which we choose to be either a step function $\lambda_{D,s}(z)$ as defined in Eq.~\eqref{eq:steplambda} or a power law function $\lambda_{D,p}(z)$ as defined in Eq.~\eqref{eq:powerlambda}. The reaction rates calculated with these dynamical $\lambda_{D}(z)$ couplings are used to produce the results presented in the main text.

\bibliographystyle{apsrev4-1}
\bibliography{reference}

\begin{thebibliography}{55}%
\makeatletter
\providecommand \@ifxundefined [1]{%
 \@ifx{#1\undefined}
}%
\providecommand \@ifnum [1]{%
 \ifnum #1\expandafter \@firstoftwo
 \else \expandafter \@secondoftwo
 \fi
}%
\providecommand \@ifx [1]{%
 \ifx #1\expandafter \@firstoftwo
 \else \expandafter \@secondoftwo
 \fi
}%
\providecommand \natexlab [1]{#1}%
\providecommand \enquote  [1]{``#1''}%
\providecommand \bibnamefont  [1]{#1}%
\providecommand \bibfnamefont [1]{#1}%
\providecommand \citenamefont [1]{#1}%
\providecommand \href@noop [0]{\@secondoftwo}%
\providecommand \href [0]{\begingroup \@sanitize@url \@href}%
\providecommand \@href[1]{\@@startlink{#1}\@@href}%
\providecommand \@@href[1]{\endgroup#1\@@endlink}%
\providecommand \@sanitize@url [0]{\catcode `\\12\catcode `\$12\catcode
  `\&12\catcode `\#12\catcode `\^12\catcode `\_12\catcode `\%12\relax}%
\providecommand \@@startlink[1]{}%
\providecommand \@@endlink[0]{}%
\providecommand \url  [0]{\begingroup\@sanitize@url \@url }%
\providecommand \@url [1]{\endgroup\@href {#1}{\urlprefix }}%
\providecommand \urlprefix  [0]{URL }%
\providecommand \Eprint [0]{\href }%
\providecommand \doibase [0]{http://dx.doi.org/}%
\providecommand \selectlanguage [0]{\@gobble}%
\providecommand \bibinfo  [0]{\@secondoftwo}%
\providecommand \bibfield  [0]{\@secondoftwo}%
\providecommand \translation [1]{[#1]}%
\providecommand \BibitemOpen [0]{}%
\providecommand \bibitemStop [0]{}%
\providecommand \bibitemNoStop [0]{.\EOS\space}%
\providecommand \EOS [0]{\spacefactor3000\relax}%
\providecommand \BibitemShut  [1]{\csname bibitem#1\endcsname}%
\let\auto@bib@innerbib\@empty
\bibitem [{\citenamefont {Fukugita}\ and\ \citenamefont
  {Yanagida}(1986)}]{Fukugita:1986hr}%
  \BibitemOpen
  \bibfield  {author} {\bibinfo {author} {\bibfnamefont {M.}~\bibnamefont
  {Fukugita}}\ and\ \bibinfo {author} {\bibfnamefont {T.}~\bibnamefont
  {Yanagida}},\ }\href {\doibase 10.1016/0370-2693(86)91126-3} {\bibfield
  {journal} {\bibinfo  {journal} {Phys. Lett. B}\ }\textbf {\bibinfo {volume}
  {174}},\ \bibinfo {pages} {45} (\bibinfo {year} {1986})}\BibitemShut
  {NoStop}%
\bibitem [{\citenamefont {Luty}(1992)}]{Luty:1992un}%
  \BibitemOpen
  \bibfield  {author} {\bibinfo {author} {\bibfnamefont {M.~A.}\ \bibnamefont
  {Luty}},\ }\href {\doibase 10.1103/PhysRevD.45.455} {\bibfield  {journal}
  {\bibinfo  {journal} {Phys. Rev. D}\ }\textbf {\bibinfo {volume} {45}},\
  \bibinfo {pages} {455} (\bibinfo {year} {1992})}\BibitemShut {NoStop}%
\bibitem [{\citenamefont {Davidson}\ \emph {et~al.}(2008)\citenamefont
  {Davidson}, \citenamefont {Nardi},\ and\ \citenamefont
  {Nir}}]{Davidson:2008bu}%
  \BibitemOpen
  \bibfield  {author} {\bibinfo {author} {\bibfnamefont {S.}~\bibnamefont
  {Davidson}}, \bibinfo {author} {\bibfnamefont {E.}~\bibnamefont {Nardi}}, \
  and\ \bibinfo {author} {\bibfnamefont {Y.}~\bibnamefont {Nir}},\ }\href
  {\doibase 10.1016/j.physrep.2008.06.002} {\bibfield  {journal} {\bibinfo
  {journal} {Phys. Rept.}\ }\textbf {\bibinfo {volume} {466}},\ \bibinfo
  {pages} {105} (\bibinfo {year} {2008})}\BibitemShut {NoStop}%
\bibitem [{\citenamefont {Plumacher}(1998)}]{Plumacher:1997ru}%
  \BibitemOpen
  \bibfield  {author} {\bibinfo {author} {\bibfnamefont {M.}~\bibnamefont
  {Plumacher}},\ }\href {\doibase 10.1016/S0550-3213(98)00410-6} {\bibfield
  {journal} {\bibinfo  {journal} {Nucl. Phys. B}\ }\textbf {\bibinfo {volume}
  {530}},\ \bibinfo {pages} {207} (\bibinfo {year} {1998})}\BibitemShut
  {NoStop}%
\bibitem [{\citenamefont {Buchmuller}\ \emph {et~al.}(2002)\citenamefont
  {Buchmuller}, \citenamefont {Di~Bari},\ and\ \citenamefont
  {Plumacher}}]{Buchmuller:2002rq}%
  \BibitemOpen
  \bibfield  {author} {\bibinfo {author} {\bibfnamefont {W.}~\bibnamefont
  {Buchmuller}}, \bibinfo {author} {\bibfnamefont {P.}~\bibnamefont {Di~Bari}},
  \ and\ \bibinfo {author} {\bibfnamefont {M.}~\bibnamefont {Plumacher}},\
  }\href {\doibase 10.1016/S0550-3213(02)00737-X} {\bibfield  {journal}
  {\bibinfo  {journal} {Nucl. Phys. B}\ }\textbf {\bibinfo {volume} {643}},\
  \bibinfo {pages} {367} (\bibinfo {year} {2002})},\ \bibinfo {note} {[Erratum:
  Nucl.Phys.B 793, 362 (2008)]}\BibitemShut {NoStop}%
\bibitem [{\citenamefont {Davis}(1994)}]{Davis:1994jw}%
  \BibitemOpen
  \bibfield  {author} {\bibinfo {author} {\bibfnamefont {R.}~\bibnamefont
  {Davis}},\ }\href {\doibase 10.1016/0146-6410(94)90004-3} {\bibfield
  {journal} {\bibinfo  {journal} {Prog. Part. Nucl. Phys.}\ }\textbf {\bibinfo
  {volume} {32}},\ \bibinfo {pages} {13} (\bibinfo {year} {1994})}\BibitemShut
  {NoStop}%
\bibitem [{\citenamefont {Fukuda}\ \emph {et~al.}(1998)\citenamefont {Fukuda}
  \emph {et~al.}}]{Super-Kamiokande:1998kpq}%
  \BibitemOpen
  \bibfield  {author} {\bibinfo {author} {\bibfnamefont {Y.}~\bibnamefont
  {Fukuda}} \emph {et~al.} (\bibinfo {collaboration} {Super-Kamiokande}),\
  }\href {\doibase 10.1103/PhysRevLett.81.1562} {\bibfield  {journal} {\bibinfo
   {journal} {Phys. Rev. Lett.}\ }\textbf {\bibinfo {volume} {81}},\ \bibinfo
  {pages} {1562} (\bibinfo {year} {1998})}\BibitemShut {NoStop}%
\bibitem [{\citenamefont {Eguchi}\ \emph {et~al.}(2003)\citenamefont {Eguchi}
  \emph {et~al.}}]{KamLAND:2002uet}%
  \BibitemOpen
  \bibfield  {author} {\bibinfo {author} {\bibfnamefont {K.}~\bibnamefont
  {Eguchi}} \emph {et~al.} (\bibinfo {collaboration} {KamLAND}),\ }\href
  {\doibase 10.1103/PhysRevLett.90.021802} {\bibfield  {journal} {\bibinfo
  {journal} {Phys. Rev. Lett.}\ }\textbf {\bibinfo {volume} {90}},\ \bibinfo
  {pages} {021802} (\bibinfo {year} {2003})}\BibitemShut {NoStop}%
\bibitem [{\citenamefont {Minkowski}(1977)}]{Minkowski:1977sc}%
  \BibitemOpen
  \bibfield  {author} {\bibinfo {author} {\bibfnamefont {P.}~\bibnamefont
  {Minkowski}},\ }\href {\doibase 10.1016/0370-2693(77)90435-X} {\bibfield
  {journal} {\bibinfo  {journal} {Phys. Lett. B}\ }\textbf {\bibinfo {volume}
  {67}},\ \bibinfo {pages} {421} (\bibinfo {year} {1977})}\BibitemShut
  {NoStop}%
\bibitem [{\citenamefont {Buchmuller}\ \emph {et~al.}(2005)\citenamefont
  {Buchmuller}, \citenamefont {Di~Bari},\ and\ \citenamefont
  {Plumacher}}]{Buchmuller:2004nz}%
  \BibitemOpen
  \bibfield  {author} {\bibinfo {author} {\bibfnamefont {W.}~\bibnamefont
  {Buchmuller}}, \bibinfo {author} {\bibfnamefont {P.}~\bibnamefont {Di~Bari}},
  \ and\ \bibinfo {author} {\bibfnamefont {M.}~\bibnamefont {Plumacher}},\
  }\href {\doibase 10.1016/j.aop.2004.02.003} {\bibfield  {journal} {\bibinfo
  {journal} {Annals Phys.}\ }\textbf {\bibinfo {volume} {315}},\ \bibinfo
  {pages} {305} (\bibinfo {year} {2005})}\BibitemShut {NoStop}%
\bibitem [{\citenamefont {Davidson}\ and\ \citenamefont
  {Ibarra}(2002)}]{Davidson:2002qv}%
  \BibitemOpen
  \bibfield  {author} {\bibinfo {author} {\bibfnamefont {S.}~\bibnamefont
  {Davidson}}\ and\ \bibinfo {author} {\bibfnamefont {A.}~\bibnamefont
  {Ibarra}},\ }\href {\doibase 10.1016/S0370-2693(02)01735-5} {\bibfield
  {journal} {\bibinfo  {journal} {Phys. Lett. B}\ }\textbf {\bibinfo {volume}
  {535}},\ \bibinfo {pages} {25} (\bibinfo {year} {2002})}\BibitemShut
  {NoStop}%
\bibitem [{\citenamefont {Cohen}\ \emph {et~al.}(2008)\citenamefont {Cohen},
  \citenamefont {Morrissey},\ and\ \citenamefont {Pierce}}]{Cohen:2008nb}%
  \BibitemOpen
  \bibfield  {author} {\bibinfo {author} {\bibfnamefont {T.}~\bibnamefont
  {Cohen}}, \bibinfo {author} {\bibfnamefont {D.~E.}\ \bibnamefont
  {Morrissey}}, \ and\ \bibinfo {author} {\bibfnamefont {A.}~\bibnamefont
  {Pierce}},\ }\href {\doibase 10.1103/PhysRevD.78.111701} {\bibfield
  {journal} {\bibinfo  {journal} {Phys. Rev. D}\ }\textbf {\bibinfo {volume}
  {78}},\ \bibinfo {pages} {111701} (\bibinfo {year} {2008})}\BibitemShut
  {NoStop}%
\bibitem [{\citenamefont {Cui}\ \emph {et~al.}(2011)\citenamefont {Cui},
  \citenamefont {Randall},\ and\ \citenamefont {Shuve}}]{Cui:2011qe}%
  \BibitemOpen
  \bibfield  {author} {\bibinfo {author} {\bibfnamefont {Y.}~\bibnamefont
  {Cui}}, \bibinfo {author} {\bibfnamefont {L.}~\bibnamefont {Randall}}, \ and\
  \bibinfo {author} {\bibfnamefont {B.}~\bibnamefont {Shuve}},\ }\href
  {\doibase 10.1007/JHEP08(2011)073} {\bibfield  {journal} {\bibinfo  {journal}
  {JHEP}\ }\textbf {\bibinfo {volume} {08}},\ \bibinfo {pages} {073} (\bibinfo
  {year} {2011})}\BibitemShut {NoStop}%
\bibitem [{\citenamefont {Baker}\ and\ \citenamefont
  {Kopp}(2017)}]{Baker:2016xzo}%
  \BibitemOpen
  \bibfield  {author} {\bibinfo {author} {\bibfnamefont {M.~J.}\ \bibnamefont
  {Baker}}\ and\ \bibinfo {author} {\bibfnamefont {J.}~\bibnamefont {Kopp}},\
  }\href {\doibase 10.1103/PhysRevLett.119.061801} {\bibfield  {journal}
  {\bibinfo  {journal} {Phys. Rev. Lett.}\ }\textbf {\bibinfo {volume} {119}},\
  \bibinfo {pages} {061801} (\bibinfo {year} {2017})}\BibitemShut {NoStop}%
\bibitem [{\citenamefont {Baker}\ \emph {et~al.}(2018)\citenamefont {Baker},
  \citenamefont {Breitbach}, \citenamefont {Kopp},\ and\ \citenamefont
  {Mittnacht}}]{Baker:2017zwx}%
  \BibitemOpen
  \bibfield  {author} {\bibinfo {author} {\bibfnamefont {M.~J.}\ \bibnamefont
  {Baker}}, \bibinfo {author} {\bibfnamefont {M.}~\bibnamefont {Breitbach}},
  \bibinfo {author} {\bibfnamefont {J.}~\bibnamefont {Kopp}}, \ and\ \bibinfo
  {author} {\bibfnamefont {L.}~\bibnamefont {Mittnacht}},\ }\href {\doibase
  10.1007/JHEP03(2018)114} {\bibfield  {journal} {\bibinfo  {journal} {JHEP}\
  }\textbf {\bibinfo {volume} {03}},\ \bibinfo {pages} {114} (\bibinfo {year}
  {2018})}\BibitemShut {NoStop}%
\bibitem [{\citenamefont {Croon}\ \emph {et~al.}(2020)\citenamefont {Croon},
  \citenamefont {Howard}, \citenamefont {Ipek},\ and\ \citenamefont
  {Tait}}]{Croon:2019ugf}%
  \BibitemOpen
  \bibfield  {author} {\bibinfo {author} {\bibfnamefont {D.}~\bibnamefont
  {Croon}}, \bibinfo {author} {\bibfnamefont {J.~N.}\ \bibnamefont {Howard}},
  \bibinfo {author} {\bibfnamefont {S.}~\bibnamefont {Ipek}}, \ and\ \bibinfo
  {author} {\bibfnamefont {T.~M.~P.}\ \bibnamefont {Tait}},\ }\href {\doibase
  10.1103/PhysRevD.101.055042} {\bibfield  {journal} {\bibinfo  {journal}
  {Phys. Rev. D}\ }\textbf {\bibinfo {volume} {101}},\ \bibinfo {pages}
  {055042} (\bibinfo {year} {2020})}\BibitemShut {NoStop}%
\bibitem [{\citenamefont {Heurtier}\ and\ \citenamefont
  {Partouche}(2020)}]{Heurtier:2019beu}%
  \BibitemOpen
  \bibfield  {author} {\bibinfo {author} {\bibfnamefont {L.}~\bibnamefont
  {Heurtier}}\ and\ \bibinfo {author} {\bibfnamefont {H.}~\bibnamefont
  {Partouche}},\ }\href {\doibase 10.1103/PhysRevD.101.043527} {\bibfield
  {journal} {\bibinfo  {journal} {Phys. Rev. D}\ }\textbf {\bibinfo {volume}
  {101}},\ \bibinfo {pages} {043527} (\bibinfo {year} {2020})}\BibitemShut
  {NoStop}%
\bibitem [{\citenamefont {Croon}\ \emph
  {et~al.}(2022{\natexlab{a}})\citenamefont {Croon}, \citenamefont {Elor},
  \citenamefont {Houtz}, \citenamefont {Murayama},\ and\ \citenamefont
  {White}}]{Croon:2020ntf}%
  \BibitemOpen
  \bibfield  {author} {\bibinfo {author} {\bibfnamefont {D.}~\bibnamefont
  {Croon}}, \bibinfo {author} {\bibfnamefont {G.}~\bibnamefont {Elor}},
  \bibinfo {author} {\bibfnamefont {R.}~\bibnamefont {Houtz}}, \bibinfo
  {author} {\bibfnamefont {H.}~\bibnamefont {Murayama}}, \ and\ \bibinfo
  {author} {\bibfnamefont {G.}~\bibnamefont {White}},\ }\href {\doibase
  10.1103/PhysRevD.105.L061303} {\bibfield  {journal} {\bibinfo  {journal}
  {Phys. Rev. D}\ }\textbf {\bibinfo {volume} {105}},\ \bibinfo {pages}
  {L061303} (\bibinfo {year} {2022}{\natexlab{a}})}\BibitemShut {NoStop}%
\bibitem [{\citenamefont {Dutta}\ \emph {et~al.}(2018)\citenamefont {Dutta},
  \citenamefont {Fong}, \citenamefont {Jimenez},\ and\ \citenamefont
  {Nardi}}]{Dutta:2018zkg}%
  \BibitemOpen
  \bibfield  {author} {\bibinfo {author} {\bibfnamefont {B.}~\bibnamefont
  {Dutta}}, \bibinfo {author} {\bibfnamefont {C.~S.}\ \bibnamefont {Fong}},
  \bibinfo {author} {\bibfnamefont {E.}~\bibnamefont {Jimenez}}, \ and\
  \bibinfo {author} {\bibfnamefont {E.}~\bibnamefont {Nardi}},\ }\href
  {\doibase 10.1088/1475-7516/2018/10/025} {\bibfield  {journal} {\bibinfo
  {journal} {JCAP}\ }\textbf {\bibinfo {volume} {10}},\ \bibinfo {pages} {025}
  (\bibinfo {year} {2018})}\BibitemShut {NoStop}%
\bibitem [{\citenamefont {Ellis}\ \emph {et~al.}(2019)\citenamefont {Ellis},
  \citenamefont {Ipek},\ and\ \citenamefont {White}}]{Ellis:2019flb}%
  \BibitemOpen
  \bibfield  {author} {\bibinfo {author} {\bibfnamefont {S.~A.~R.}\
  \bibnamefont {Ellis}}, \bibinfo {author} {\bibfnamefont {S.}~\bibnamefont
  {Ipek}}, \ and\ \bibinfo {author} {\bibfnamefont {G.}~\bibnamefont {White}},\
  }\href {\doibase 10.1007/JHEP08(2019)002} {\bibfield  {journal} {\bibinfo
  {journal} {JHEP}\ }\textbf {\bibinfo {volume} {08}},\ \bibinfo {pages} {002}
  (\bibinfo {year} {2019})}\BibitemShut {NoStop}%
\bibitem [{\citenamefont {Azatov}\ \emph {et~al.}(2021)\citenamefont {Azatov},
  \citenamefont {Vanvlasselaer},\ and\ \citenamefont {Yin}}]{Azatov:2021irb}%
  \BibitemOpen
  \bibfield  {author} {\bibinfo {author} {\bibfnamefont {A.}~\bibnamefont
  {Azatov}}, \bibinfo {author} {\bibfnamefont {M.}~\bibnamefont
  {Vanvlasselaer}}, \ and\ \bibinfo {author} {\bibfnamefont {W.}~\bibnamefont
  {Yin}},\ }\href {\doibase 10.1007/JHEP10(2021)043} {\bibfield  {journal}
  {\bibinfo  {journal} {JHEP}\ }\textbf {\bibinfo {volume} {10}},\ \bibinfo
  {pages} {043} (\bibinfo {year} {2021})}\BibitemShut {NoStop}%
\bibitem [{\citenamefont {Baldes}\ \emph {et~al.}(2021)\citenamefont {Baldes},
  \citenamefont {Blasi}, \citenamefont {Mariotti}, \citenamefont {Sevrin},\
  and\ \citenamefont {Turbang}}]{Baldes:2021vyz}%
  \BibitemOpen
  \bibfield  {author} {\bibinfo {author} {\bibfnamefont {I.}~\bibnamefont
  {Baldes}}, \bibinfo {author} {\bibfnamefont {S.}~\bibnamefont {Blasi}},
  \bibinfo {author} {\bibfnamefont {A.}~\bibnamefont {Mariotti}}, \bibinfo
  {author} {\bibfnamefont {A.}~\bibnamefont {Sevrin}}, \ and\ \bibinfo {author}
  {\bibfnamefont {K.}~\bibnamefont {Turbang}},\ }\href {\doibase
  10.1103/PhysRevD.104.115029} {\bibfield  {journal} {\bibinfo  {journal}
  {Phys. Rev. D}\ }\textbf {\bibinfo {volume} {104}},\ \bibinfo {pages}
  {115029} (\bibinfo {year} {2021})}\BibitemShut {NoStop}%
\bibitem [{\citenamefont {Croon}\ \emph
  {et~al.}(2022{\natexlab{b}})\citenamefont {Croon}, \citenamefont
  {Davoudiasl},\ and\ \citenamefont {Houtz}}]{Croon:2022gwq}%
  \BibitemOpen
  \bibfield  {author} {\bibinfo {author} {\bibfnamefont {D.}~\bibnamefont
  {Croon}}, \bibinfo {author} {\bibfnamefont {H.}~\bibnamefont {Davoudiasl}}, \
  and\ \bibinfo {author} {\bibfnamefont {R.}~\bibnamefont {Houtz}},\ }\href
  {\doibase 10.1103/PhysRevD.106.035006} {\bibfield  {journal} {\bibinfo
  {journal} {Phys. Rev. D}\ }\textbf {\bibinfo {volume} {106}},\ \bibinfo
  {pages} {035006} (\bibinfo {year} {2022}{\natexlab{b}})}\BibitemShut
  {NoStop}%
\bibitem [{\citenamefont {Huang}\ and\ \citenamefont
  {Xie}(2022)}]{Huang:2022vkf}%
  \BibitemOpen
  \bibfield  {author} {\bibinfo {author} {\bibfnamefont {P.}~\bibnamefont
  {Huang}}\ and\ \bibinfo {author} {\bibfnamefont {K.-P.}\ \bibnamefont
  {Xie}},\ }\href {\doibase 10.1007/JHEP09(2022)052} {\bibfield  {journal}
  {\bibinfo  {journal} {JHEP}\ }\textbf {\bibinfo {volume} {09}},\ \bibinfo
  {pages} {052} (\bibinfo {year} {2022})}\BibitemShut {NoStop}%
\bibitem [{\citenamefont {Chun}\ \emph {et~al.}(2023)\citenamefont {Chun},
  \citenamefont {Dutka}, \citenamefont {Jung}, \citenamefont {Nagels},\ and\
  \citenamefont {Vanvlasselaer}}]{Chun:2023ezg}%
  \BibitemOpen
  \bibfield  {author} {\bibinfo {author} {\bibfnamefont {E.~J.}\ \bibnamefont
  {Chun}}, \bibinfo {author} {\bibfnamefont {T.~P.}\ \bibnamefont {Dutka}},
  \bibinfo {author} {\bibfnamefont {T.~H.}\ \bibnamefont {Jung}}, \bibinfo
  {author} {\bibfnamefont {X.}~\bibnamefont {Nagels}}, \ and\ \bibinfo {author}
  {\bibfnamefont {M.}~\bibnamefont {Vanvlasselaer}},\ }\href {\doibase
  10.1007/JHEP09(2023)164} {\bibfield  {journal} {\bibinfo  {journal} {JHEP}\
  }\textbf {\bibinfo {volume} {09}},\ \bibinfo {pages} {164} (\bibinfo {year}
  {2023})}\BibitemShut {NoStop}%
\bibitem [{\citenamefont {ChoeJo}\ \emph
  {et~al.}(2023{\natexlab{a}})\citenamefont {ChoeJo}, \citenamefont {Kim},\
  and\ \citenamefont {Lee}}]{ChoeJo:2023ffp}%
  \BibitemOpen
  \bibfield  {author} {\bibinfo {author} {\bibfnamefont {Y.}~\bibnamefont
  {ChoeJo}}, \bibinfo {author} {\bibfnamefont {Y.}~\bibnamefont {Kim}}, \ and\
  \bibinfo {author} {\bibfnamefont {H.-S.}\ \bibnamefont {Lee}},\ }\href
  {\doibase 10.1103/PhysRevD.108.095028} {\bibfield  {journal} {\bibinfo
  {journal} {Phys. Rev. D}\ }\textbf {\bibinfo {volume} {108}},\ \bibinfo
  {pages} {095028} (\bibinfo {year} {2023}{\natexlab{a}})}\BibitemShut
  {NoStop}%
\bibitem [{\citenamefont {ChoeJo}\ \emph
  {et~al.}(2023{\natexlab{b}})\citenamefont {ChoeJo}, \citenamefont {Enomoto},
  \citenamefont {Kim},\ and\ \citenamefont {Lee}}]{ChoeJo:2023cnx}%
  \BibitemOpen
  \bibfield  {author} {\bibinfo {author} {\bibfnamefont {Y.}~\bibnamefont
  {ChoeJo}}, \bibinfo {author} {\bibfnamefont {K.}~\bibnamefont {Enomoto}},
  \bibinfo {author} {\bibfnamefont {Y.}~\bibnamefont {Kim}}, \ and\ \bibinfo
  {author} {\bibfnamefont {H.-S.}\ \bibnamefont {Lee}},\ }\href@noop {} {\
  (\bibinfo {year} {2023}{\natexlab{b}})},\ \Eprint
  {http://arxiv.org/abs/2311.16672} {arXiv:2311.16672 [hep-ph]} \BibitemShut
  {NoStop}%
\bibitem [{\citenamefont {Abada}\ \emph
  {et~al.}(2006{\natexlab{a}})\citenamefont {Abada}, \citenamefont {Davidson},
  \citenamefont {Ibarra}, \citenamefont {Josse-Michaux}, \citenamefont
  {Losada},\ and\ \citenamefont {Riotto}}]{Abada:2006ea}%
  \BibitemOpen
  \bibfield  {author} {\bibinfo {author} {\bibfnamefont {A.}~\bibnamefont
  {Abada}}, \bibinfo {author} {\bibfnamefont {S.}~\bibnamefont {Davidson}},
  \bibinfo {author} {\bibfnamefont {A.}~\bibnamefont {Ibarra}}, \bibinfo
  {author} {\bibfnamefont {F.~X.}\ \bibnamefont {Josse-Michaux}}, \bibinfo
  {author} {\bibfnamefont {M.}~\bibnamefont {Losada}}, \ and\ \bibinfo {author}
  {\bibfnamefont {A.}~\bibnamefont {Riotto}},\ }\href {\doibase
  10.1088/1126-6708/2006/09/010} {\bibfield  {journal} {\bibinfo  {journal}
  {JHEP}\ }\textbf {\bibinfo {volume} {09}},\ \bibinfo {pages} {010} (\bibinfo
  {year} {2006}{\natexlab{a}})},\ \Eprint {http://arxiv.org/abs/hep-ph/0605281}
  {arXiv:hep-ph/0605281} \BibitemShut {NoStop}%
\bibitem [{\citenamefont {Buchmuller}\ \emph {et~al.}(2003)\citenamefont
  {Buchmuller}, \citenamefont {Di~Bari},\ and\ \citenamefont
  {Plumacher}}]{Buchmuller:2003gz}%
  \BibitemOpen
  \bibfield  {author} {\bibinfo {author} {\bibfnamefont {W.}~\bibnamefont
  {Buchmuller}}, \bibinfo {author} {\bibfnamefont {P.}~\bibnamefont {Di~Bari}},
  \ and\ \bibinfo {author} {\bibfnamefont {M.}~\bibnamefont {Plumacher}},\
  }\href {\doibase 10.1016/S0550-3213(03)00449-8} {\bibfield  {journal}
  {\bibinfo  {journal} {Nucl. Phys. B}\ }\textbf {\bibinfo {volume} {665}},\
  \bibinfo {pages} {445} (\bibinfo {year} {2003})},\ \Eprint
  {http://arxiv.org/abs/hep-ph/0302092} {arXiv:hep-ph/0302092} \BibitemShut
  {NoStop}%
\bibitem [{\citenamefont {Workman}\ and\ \citenamefont
  {Others}(2022)}]{Workman:2022ynf}%
  \BibitemOpen
  \bibfield  {author} {\bibinfo {author} {\bibfnamefont {R.~L.}\ \bibnamefont
  {Workman}}\ and\ \bibinfo {author} {\bibnamefont {Others}} (\bibinfo
  {collaboration} {Particle Data Group}),\ }\href {\doibase
  10.1093/ptep/ptac097} {\bibfield  {journal} {\bibinfo  {journal} {PTEP}\
  }\textbf {\bibinfo {volume} {2022}},\ \bibinfo {pages} {083C01} (\bibinfo
  {year} {2022})}\BibitemShut {NoStop}%
\bibitem [{\citenamefont {Barbieri}\ \emph {et~al.}(2000)\citenamefont
  {Barbieri}, \citenamefont {Creminelli}, \citenamefont {Strumia},\ and\
  \citenamefont {Tetradis}}]{Barbieri:1999ma}%
  \BibitemOpen
  \bibfield  {author} {\bibinfo {author} {\bibfnamefont {R.}~\bibnamefont
  {Barbieri}}, \bibinfo {author} {\bibfnamefont {P.}~\bibnamefont
  {Creminelli}}, \bibinfo {author} {\bibfnamefont {A.}~\bibnamefont {Strumia}},
  \ and\ \bibinfo {author} {\bibfnamefont {N.}~\bibnamefont {Tetradis}},\
  }\href {\doibase 10.1016/S0550-3213(00)00011-0} {\bibfield  {journal}
  {\bibinfo  {journal} {Nucl. Phys. B}\ }\textbf {\bibinfo {volume} {575}},\
  \bibinfo {pages} {61} (\bibinfo {year} {2000})},\ \Eprint
  {http://arxiv.org/abs/hep-ph/9911315} {arXiv:hep-ph/9911315} \BibitemShut
  {NoStop}%
\bibitem [{\citenamefont {Graham}\ \emph {et~al.}(2015)\citenamefont {Graham},
  \citenamefont {Kaplan},\ and\ \citenamefont {Rajendran}}]{Graham:2015cka}%
  \BibitemOpen
  \bibfield  {author} {\bibinfo {author} {\bibfnamefont {P.~W.}\ \bibnamefont
  {Graham}}, \bibinfo {author} {\bibfnamefont {D.~E.}\ \bibnamefont {Kaplan}},
  \ and\ \bibinfo {author} {\bibfnamefont {S.}~\bibnamefont {Rajendran}},\
  }\href {\doibase 10.1103/PhysRevLett.115.221801} {\bibfield  {journal}
  {\bibinfo  {journal} {Phys. Rev. Lett.}\ }\textbf {\bibinfo {volume} {115}},\
  \bibinfo {pages} {221801} (\bibinfo {year} {2015})}\BibitemShut {NoStop}%
\bibitem [{\citenamefont {Kawamura}\ \emph {et~al.}(2011)\citenamefont
  {Kawamura} \emph {et~al.}}]{Decigo}%
  \BibitemOpen
  \bibfield  {author} {\bibinfo {author} {\bibfnamefont {S.}~\bibnamefont
  {Kawamura}} \emph {et~al.},\ }\href {\doibase 10.1088/0264-9381/28/9/094011}
  {\bibfield  {journal} {\bibinfo  {journal} {Classical and Quantum Gravity}\
  }\textbf {\bibinfo {volume} {28}},\ \bibinfo {pages} {094011} (\bibinfo
  {year} {2011})}\BibitemShut {NoStop}%
\bibitem [{\citenamefont {Crowder}\ and\ \citenamefont
  {Cornish}(2005)}]{BBO:2005nr}%
  \BibitemOpen
  \bibfield  {author} {\bibinfo {author} {\bibfnamefont {J.}~\bibnamefont
  {Crowder}}\ and\ \bibinfo {author} {\bibfnamefont {N.~J.}\ \bibnamefont
  {Cornish}},\ }\href {\doibase 10.1103/PhysRevD.72.083005} {\bibfield
  {journal} {\bibinfo  {journal} {Phys. Rev. D}\ }\textbf {\bibinfo {volume}
  {72}},\ \bibinfo {pages} {083005} (\bibinfo {year} {2005})},\ \Eprint
  {http://arxiv.org/abs/gr-qc/0506015} {arXiv:gr-qc/0506015} \BibitemShut
  {NoStop}%
\bibitem [{\citenamefont {Abbott}\ \emph {et~al.}(2019)\citenamefont {Abbott}
  \emph {et~al.}}]{aLIGO:2019vic}%
  \BibitemOpen
  \bibfield  {author} {\bibinfo {author} {\bibfnamefont {B.~P.}\ \bibnamefont
  {Abbott}} \emph {et~al.} (\bibinfo {collaboration} {LIGO Scientific,
  Virgo}),\ }\href {\doibase 10.1103/PhysRevD.100.061101} {\bibfield  {journal}
  {\bibinfo  {journal} {Phys. Rev. D}\ }\textbf {\bibinfo {volume} {100}},\
  \bibinfo {pages} {061101} (\bibinfo {year} {2019})},\ \Eprint
  {http://arxiv.org/abs/1903.02886} {arXiv:1903.02886 [gr-qc]} \BibitemShut
  {NoStop}%
\bibitem [{\citenamefont {Hild}\ \emph {et~al.}(2011)\citenamefont {Hild} \emph
  {et~al.}}]{ET:2010id}%
  \BibitemOpen
  \bibfield  {author} {\bibinfo {author} {\bibfnamefont {S.}~\bibnamefont
  {Hild}} \emph {et~al.},\ }\href {\doibase 10.1088/0264-9381/28/9/094013}
  {\bibfield  {journal} {\bibinfo  {journal} {Class. Quant. Grav.}\ }\textbf
  {\bibinfo {volume} {28}},\ \bibinfo {pages} {094013} (\bibinfo {year}
  {2011})},\ \Eprint {http://arxiv.org/abs/1012.0908} {arXiv:1012.0908 [gr-qc]}
  \BibitemShut {NoStop}%
\bibitem [{\citenamefont {Sathyaprakash}\ \emph {et~al.}(2012)\citenamefont
  {Sathyaprakash} \emph {et~al.}}]{ET:2012jk}%
  \BibitemOpen
  \bibfield  {author} {\bibinfo {author} {\bibfnamefont {B.}~\bibnamefont
  {Sathyaprakash}} \emph {et~al.},\ }\href {\doibase
  10.1088/0264-9381/29/12/124013} {\bibfield  {journal} {\bibinfo  {journal}
  {Class. Quant. Grav.}\ }\textbf {\bibinfo {volume} {29}},\ \bibinfo {pages}
  {124013} (\bibinfo {year} {2012})},\ \bibinfo {note} {[Erratum:
  Class.Quant.Grav. 30, 079501 (2013)]},\ \Eprint
  {http://arxiv.org/abs/1206.0331} {arXiv:1206.0331 [gr-qc]} \BibitemShut
  {NoStop}%
\bibitem [{\citenamefont {{Reitze}}\ \emph {et~al.}(2019)\citenamefont
  {{Reitze}}, \citenamefont {{Adhikari}}, \citenamefont {{Ballmer}},
  \citenamefont {{Barish}}, \citenamefont {{Barsotti}}, \citenamefont
  {{Billingsley}}, \citenamefont {{Brown}}, \citenamefont {{Chen}},
  \citenamefont {{Coyne}}, \citenamefont {{Eisenstein}}, \citenamefont
  {{Evans}}, \citenamefont {{Fritschel}}, \citenamefont {{Hall}}, \citenamefont
  {{Lazzarini}}, \citenamefont {{Lovelace}}, \citenamefont {{Read}},
  \citenamefont {{Sathyaprakash}}, \citenamefont {{Shoemaker}}, \citenamefont
  {{Smith}}, \citenamefont {{Torrie}}, \citenamefont {{Vitale}}, \citenamefont
  {{Weiss}}, \citenamefont {{Wipf}},\ and\ \citenamefont
  {{Zucker}}}]{2019BAAS...51g..35R}%
  \BibitemOpen
  \bibfield  {author} {\bibinfo {author} {\bibfnamefont {D.}~\bibnamefont
  {{Reitze}}}, \bibinfo {author} {\bibfnamefont {R.~X.}\ \bibnamefont
  {{Adhikari}}}, \bibinfo {author} {\bibfnamefont {S.}~\bibnamefont
  {{Ballmer}}}, \bibinfo {author} {\bibfnamefont {B.}~\bibnamefont {{Barish}}},
  \bibinfo {author} {\bibfnamefont {L.}~\bibnamefont {{Barsotti}}}, \bibinfo
  {author} {\bibfnamefont {G.}~\bibnamefont {{Billingsley}}}, \bibinfo {author}
  {\bibfnamefont {D.~A.}\ \bibnamefont {{Brown}}}, \bibinfo {author}
  {\bibfnamefont {Y.}~\bibnamefont {{Chen}}}, \bibinfo {author} {\bibfnamefont
  {D.}~\bibnamefont {{Coyne}}}, \bibinfo {author} {\bibfnamefont
  {R.}~\bibnamefont {{Eisenstein}}}, \bibinfo {author} {\bibfnamefont
  {M.}~\bibnamefont {{Evans}}}, \bibinfo {author} {\bibfnamefont
  {P.}~\bibnamefont {{Fritschel}}}, \bibinfo {author} {\bibfnamefont {E.~D.}\
  \bibnamefont {{Hall}}}, \bibinfo {author} {\bibfnamefont {A.}~\bibnamefont
  {{Lazzarini}}}, \bibinfo {author} {\bibfnamefont {G.}~\bibnamefont
  {{Lovelace}}}, \bibinfo {author} {\bibfnamefont {J.}~\bibnamefont {{Read}}},
  \bibinfo {author} {\bibfnamefont {B.~S.}\ \bibnamefont {{Sathyaprakash}}},
  \bibinfo {author} {\bibfnamefont {D.}~\bibnamefont {{Shoemaker}}}, \bibinfo
  {author} {\bibfnamefont {J.}~\bibnamefont {{Smith}}}, \bibinfo {author}
  {\bibfnamefont {C.}~\bibnamefont {{Torrie}}}, \bibinfo {author}
  {\bibfnamefont {S.}~\bibnamefont {{Vitale}}}, \bibinfo {author}
  {\bibfnamefont {R.}~\bibnamefont {{Weiss}}}, \bibinfo {author} {\bibfnamefont
  {C.}~\bibnamefont {{Wipf}}}, \ and\ \bibinfo {author} {\bibfnamefont
  {M.}~\bibnamefont {{Zucker}}},\ }in\ \href {\doibase
  10.48550/arXiv.1907.04833} {\emph {\bibinfo {booktitle} {Bulletin of the
  American Astronomical Society}}},\ Vol.~\bibinfo {volume} {51}\ (\bibinfo
  {year} {2019})\ p.~\bibinfo {pages} {35},\ \Eprint
  {http://arxiv.org/abs/1907.04833} {arXiv:1907.04833 [astro-ph.IM]}
  \BibitemShut {NoStop}%
\bibitem [{\citenamefont {Franciolini}\ \emph {et~al.}(2022)\citenamefont
  {Franciolini}, \citenamefont {Maharana},\ and\ \citenamefont
  {Muia}}]{Franciolini:2022htd}%
  \BibitemOpen
  \bibfield  {author} {\bibinfo {author} {\bibfnamefont {G.}~\bibnamefont
  {Franciolini}}, \bibinfo {author} {\bibfnamefont {A.}~\bibnamefont
  {Maharana}}, \ and\ \bibinfo {author} {\bibfnamefont {F.}~\bibnamefont
  {Muia}},\ }\href {\doibase 10.1103/PhysRevD.106.103520} {\bibfield  {journal}
  {\bibinfo  {journal} {Phys. Rev. D}\ }\textbf {\bibinfo {volume} {106}},\
  \bibinfo {pages} {103520} (\bibinfo {year} {2022})},\ \Eprint
  {http://arxiv.org/abs/2205.02153} {arXiv:2205.02153 [astro-ph.CO]}
  \BibitemShut {NoStop}%
\bibitem [{\citenamefont {Ringwald}\ \emph {et~al.}(2021)\citenamefont
  {Ringwald}, \citenamefont {Sch\"utte-Engel},\ and\ \citenamefont
  {Tamarit}}]{Ringwald:2020ist}%
  \BibitemOpen
  \bibfield  {author} {\bibinfo {author} {\bibfnamefont {A.}~\bibnamefont
  {Ringwald}}, \bibinfo {author} {\bibfnamefont {J.}~\bibnamefont
  {Sch\"utte-Engel}}, \ and\ \bibinfo {author} {\bibfnamefont {C.}~\bibnamefont
  {Tamarit}},\ }\href {\doibase 10.1088/1475-7516/2021/03/054} {\bibfield
  {journal} {\bibinfo  {journal} {JCAP}\ }\textbf {\bibinfo {volume} {03}},\
  \bibinfo {pages} {054} (\bibinfo {year} {2021})},\ \Eprint
  {http://arxiv.org/abs/2011.04731} {arXiv:2011.04731 [hep-ph]} \BibitemShut
  {NoStop}%
\bibitem [{\citenamefont {Abada}\ \emph
  {et~al.}(2006{\natexlab{b}})\citenamefont {Abada}, \citenamefont {Davidson},
  \citenamefont {Josse-Michaux}, \citenamefont {Losada},\ and\ \citenamefont
  {Riotto}}]{Abada:2006fw}%
  \BibitemOpen
  \bibfield  {author} {\bibinfo {author} {\bibfnamefont {A.}~\bibnamefont
  {Abada}}, \bibinfo {author} {\bibfnamefont {S.}~\bibnamefont {Davidson}},
  \bibinfo {author} {\bibfnamefont {F.-X.}\ \bibnamefont {Josse-Michaux}},
  \bibinfo {author} {\bibfnamefont {M.}~\bibnamefont {Losada}}, \ and\ \bibinfo
  {author} {\bibfnamefont {A.}~\bibnamefont {Riotto}},\ }\href {\doibase
  10.1088/1475-7516/2006/04/004} {\bibfield  {journal} {\bibinfo  {journal}
  {JCAP}\ }\textbf {\bibinfo {volume} {04}},\ \bibinfo {pages} {004} (\bibinfo
  {year} {2006}{\natexlab{b}})},\ \Eprint {http://arxiv.org/abs/hep-ph/0601083}
  {arXiv:hep-ph/0601083} \BibitemShut {NoStop}%
\bibitem [{\citenamefont {Nardi}\ \emph {et~al.}(2006)\citenamefont {Nardi},
  \citenamefont {Nir}, \citenamefont {Roulet},\ and\ \citenamefont
  {Racker}}]{Nardi:2006fx}%
  \BibitemOpen
  \bibfield  {author} {\bibinfo {author} {\bibfnamefont {E.}~\bibnamefont
  {Nardi}}, \bibinfo {author} {\bibfnamefont {Y.}~\bibnamefont {Nir}}, \bibinfo
  {author} {\bibfnamefont {E.}~\bibnamefont {Roulet}}, \ and\ \bibinfo {author}
  {\bibfnamefont {J.}~\bibnamefont {Racker}},\ }\href {\doibase
  10.1088/1126-6708/2006/01/164} {\bibfield  {journal} {\bibinfo  {journal}
  {JHEP}\ }\textbf {\bibinfo {volume} {01}},\ \bibinfo {pages} {164} (\bibinfo
  {year} {2006})},\ \Eprint {http://arxiv.org/abs/hep-ph/0601084}
  {arXiv:hep-ph/0601084} \BibitemShut {NoStop}%
\bibitem [{\citenamefont {Blanchet}\ and\ \citenamefont
  {Di~Bari}(2007)}]{Blanchet:2006be}%
  \BibitemOpen
  \bibfield  {author} {\bibinfo {author} {\bibfnamefont {S.}~\bibnamefont
  {Blanchet}}\ and\ \bibinfo {author} {\bibfnamefont {P.}~\bibnamefont
  {Di~Bari}},\ }\href {\doibase 10.1088/1475-7516/2007/03/018} {\bibfield
  {journal} {\bibinfo  {journal} {JCAP}\ }\textbf {\bibinfo {volume} {03}},\
  \bibinfo {pages} {018} (\bibinfo {year} {2007})},\ \Eprint
  {http://arxiv.org/abs/hep-ph/0607330} {arXiv:hep-ph/0607330} \BibitemShut
  {NoStop}%
\bibitem [{\citenamefont {Pascoli}\ \emph
  {et~al.}(2007{\natexlab{a}})\citenamefont {Pascoli}, \citenamefont {Petcov},\
  and\ \citenamefont {Riotto}}]{Pascoli:2006ie}%
  \BibitemOpen
  \bibfield  {author} {\bibinfo {author} {\bibfnamefont {S.}~\bibnamefont
  {Pascoli}}, \bibinfo {author} {\bibfnamefont {S.~T.}\ \bibnamefont {Petcov}},
  \ and\ \bibinfo {author} {\bibfnamefont {A.}~\bibnamefont {Riotto}},\ }\href
  {\doibase 10.1103/PhysRevD.75.083511} {\bibfield  {journal} {\bibinfo
  {journal} {Phys. Rev. D}\ }\textbf {\bibinfo {volume} {75}},\ \bibinfo
  {pages} {083511} (\bibinfo {year} {2007}{\natexlab{a}})},\ \Eprint
  {http://arxiv.org/abs/hep-ph/0609125} {arXiv:hep-ph/0609125} \BibitemShut
  {NoStop}%
\bibitem [{\citenamefont {De~Simone}\ and\ \citenamefont
  {Riotto}(2007)}]{DeSimone:2006nrs}%
  \BibitemOpen
  \bibfield  {author} {\bibinfo {author} {\bibfnamefont {A.}~\bibnamefont
  {De~Simone}}\ and\ \bibinfo {author} {\bibfnamefont {A.}~\bibnamefont
  {Riotto}},\ }\href {\doibase 10.1088/1475-7516/2007/02/005} {\bibfield
  {journal} {\bibinfo  {journal} {JCAP}\ }\textbf {\bibinfo {volume} {02}},\
  \bibinfo {pages} {005} (\bibinfo {year} {2007})},\ \Eprint
  {http://arxiv.org/abs/hep-ph/0611357} {arXiv:hep-ph/0611357} \BibitemShut
  {NoStop}%
\bibitem [{\citenamefont {Dev}\ \emph {et~al.}(2018)\citenamefont {Dev},
  \citenamefont {Di~Bari}, \citenamefont {Garbrecht}, \citenamefont {Lavignac},
  \citenamefont {Millington},\ and\ \citenamefont {Teresi}}]{Dev:2017trv}%
  \BibitemOpen
  \bibfield  {author} {\bibinfo {author} {\bibfnamefont {P.~S.~B.}\
  \bibnamefont {Dev}}, \bibinfo {author} {\bibfnamefont {P.}~\bibnamefont
  {Di~Bari}}, \bibinfo {author} {\bibfnamefont {B.}~\bibnamefont {Garbrecht}},
  \bibinfo {author} {\bibfnamefont {S.}~\bibnamefont {Lavignac}}, \bibinfo
  {author} {\bibfnamefont {P.}~\bibnamefont {Millington}}, \ and\ \bibinfo
  {author} {\bibfnamefont {D.}~\bibnamefont {Teresi}},\ }\href {\doibase
  10.1142/S0217751X18420010} {\bibfield  {journal} {\bibinfo  {journal} {Int.
  J. Mod. Phys. A}\ }\textbf {\bibinfo {volume} {33}},\ \bibinfo {pages}
  {1842001} (\bibinfo {year} {2018})},\ \Eprint
  {http://arxiv.org/abs/1711.02861} {arXiv:1711.02861 [hep-ph]} \BibitemShut
  {NoStop}%
\bibitem [{\citenamefont {Blanchet}\ and\ \citenamefont
  {Di~Bari}(2009)}]{Blanchet:2008pw}%
  \BibitemOpen
  \bibfield  {author} {\bibinfo {author} {\bibfnamefont {S.}~\bibnamefont
  {Blanchet}}\ and\ \bibinfo {author} {\bibfnamefont {P.}~\bibnamefont
  {Di~Bari}},\ }\href {\doibase 10.1016/j.nuclphysb.2008.08.026} {\bibfield
  {journal} {\bibinfo  {journal} {Nucl. Phys. B}\ }\textbf {\bibinfo {volume}
  {807}},\ \bibinfo {pages} {155} (\bibinfo {year} {2009})},\ \Eprint
  {http://arxiv.org/abs/0807.0743} {arXiv:0807.0743 [hep-ph]} \BibitemShut
  {NoStop}%
\bibitem [{\citenamefont {Pascoli}\ \emph
  {et~al.}(2007{\natexlab{b}})\citenamefont {Pascoli}, \citenamefont {Petcov},\
  and\ \citenamefont {Riotto}}]{Pascoli:2006ci}%
  \BibitemOpen
  \bibfield  {author} {\bibinfo {author} {\bibfnamefont {S.}~\bibnamefont
  {Pascoli}}, \bibinfo {author} {\bibfnamefont {S.~T.}\ \bibnamefont {Petcov}},
  \ and\ \bibinfo {author} {\bibfnamefont {A.}~\bibnamefont {Riotto}},\ }\href
  {\doibase 10.1016/j.nuclphysb.2007.02.019} {\bibfield  {journal} {\bibinfo
  {journal} {Nucl. Phys. B}\ }\textbf {\bibinfo {volume} {774}},\ \bibinfo
  {pages} {1} (\bibinfo {year} {2007}{\natexlab{b}})},\ \Eprint
  {http://arxiv.org/abs/hep-ph/0611338} {arXiv:hep-ph/0611338} \BibitemShut
  {NoStop}%
\bibitem [{\citenamefont {Anisimov}\ \emph {et~al.}(2008)\citenamefont
  {Anisimov}, \citenamefont {Blanchet},\ and\ \citenamefont
  {Di~Bari}}]{Anisimov:2007mw}%
  \BibitemOpen
  \bibfield  {author} {\bibinfo {author} {\bibfnamefont {A.}~\bibnamefont
  {Anisimov}}, \bibinfo {author} {\bibfnamefont {S.}~\bibnamefont {Blanchet}},
  \ and\ \bibinfo {author} {\bibfnamefont {P.}~\bibnamefont {Di~Bari}},\ }\href
  {\doibase 10.1088/1475-7516/2008/04/033} {\bibfield  {journal} {\bibinfo
  {journal} {JCAP}\ }\textbf {\bibinfo {volume} {04}},\ \bibinfo {pages} {033}
  (\bibinfo {year} {2008})},\ \Eprint {http://arxiv.org/abs/0707.3024}
  {arXiv:0707.3024 [hep-ph]} \BibitemShut {NoStop}%
\bibitem [{\citenamefont {Molinaro}\ \emph {et~al.}(2008)\citenamefont
  {Molinaro}, \citenamefont {Petcov}, \citenamefont {Shindou},\ and\
  \citenamefont {Takanishi}}]{Molinaro:2007uv}%
  \BibitemOpen
  \bibfield  {author} {\bibinfo {author} {\bibfnamefont {E.}~\bibnamefont
  {Molinaro}}, \bibinfo {author} {\bibfnamefont {S.~T.}\ \bibnamefont
  {Petcov}}, \bibinfo {author} {\bibfnamefont {T.}~\bibnamefont {Shindou}}, \
  and\ \bibinfo {author} {\bibfnamefont {Y.}~\bibnamefont {Takanishi}},\ }\href
  {\doibase 10.1016/j.nuclphysb.2007.12.033} {\bibfield  {journal} {\bibinfo
  {journal} {Nucl. Phys. B}\ }\textbf {\bibinfo {volume} {797}},\ \bibinfo
  {pages} {93} (\bibinfo {year} {2008})},\ \Eprint
  {http://arxiv.org/abs/0709.0413} {arXiv:0709.0413 [hep-ph]} \BibitemShut
  {NoStop}%
\bibitem [{\citenamefont {Molinaro}\ and\ \citenamefont
  {Petcov}(2009{\natexlab{a}})}]{Molinaro:2008cw}%
  \BibitemOpen
  \bibfield  {author} {\bibinfo {author} {\bibfnamefont {E.}~\bibnamefont
  {Molinaro}}\ and\ \bibinfo {author} {\bibfnamefont {S.~T.}\ \bibnamefont
  {Petcov}},\ }\href {\doibase 10.1016/j.physletb.2008.11.047} {\bibfield
  {journal} {\bibinfo  {journal} {Phys. Lett. B}\ }\textbf {\bibinfo {volume}
  {671}},\ \bibinfo {pages} {60} (\bibinfo {year} {2009}{\natexlab{a}})},\
  \Eprint {http://arxiv.org/abs/0808.3534} {arXiv:0808.3534 [hep-ph]}
  \BibitemShut {NoStop}%
\bibitem [{\citenamefont {Molinaro}\ and\ \citenamefont
  {Petcov}(2009{\natexlab{b}})}]{Molinaro:2009lud}%
  \BibitemOpen
  \bibfield  {author} {\bibinfo {author} {\bibfnamefont {E.}~\bibnamefont
  {Molinaro}}\ and\ \bibinfo {author} {\bibfnamefont {S.~T.}\ \bibnamefont
  {Petcov}},\ }\href {\doibase 10.1140/epjc/s10052-009-0985-3} {\bibfield
  {journal} {\bibinfo  {journal} {Eur. Phys. J. C}\ }\textbf {\bibinfo {volume}
  {61}},\ \bibinfo {pages} {93} (\bibinfo {year} {2009}{\natexlab{b}})},\
  \Eprint {http://arxiv.org/abs/0803.4120} {arXiv:0803.4120 [hep-ph]}
  \BibitemShut {NoStop}%
\bibitem [{\citenamefont {Joshipura}\ \emph {et~al.}(2001)\citenamefont
  {Joshipura}, \citenamefont {Paschos},\ and\ \citenamefont
  {Rodejohann}}]{Joshipura:2001ui}%
  \BibitemOpen
  \bibfield  {author} {\bibinfo {author} {\bibfnamefont {A.~S.}\ \bibnamefont
  {Joshipura}}, \bibinfo {author} {\bibfnamefont {E.~A.}\ \bibnamefont
  {Paschos}}, \ and\ \bibinfo {author} {\bibfnamefont {W.}~\bibnamefont
  {Rodejohann}},\ }\href {\doibase 10.1088/1126-6708/2001/08/029} {\bibfield
  {journal} {\bibinfo  {journal} {JHEP}\ }\textbf {\bibinfo {volume} {08}},\
  \bibinfo {pages} {029} (\bibinfo {year} {2001})},\ \Eprint
  {http://arxiv.org/abs/hep-ph/0105175} {arXiv:hep-ph/0105175} \BibitemShut
  {NoStop}%
\bibitem [{\citenamefont {Moffat}\ \emph {et~al.}(2019)\citenamefont {Moffat},
  \citenamefont {Pascoli}, \citenamefont {Petcov},\ and\ \citenamefont
  {Turner}}]{Moffat:2018smo}%
  \BibitemOpen
  \bibfield  {author} {\bibinfo {author} {\bibfnamefont {K.}~\bibnamefont
  {Moffat}}, \bibinfo {author} {\bibfnamefont {S.}~\bibnamefont {Pascoli}},
  \bibinfo {author} {\bibfnamefont {S.~T.}\ \bibnamefont {Petcov}}, \ and\
  \bibinfo {author} {\bibfnamefont {J.}~\bibnamefont {Turner}},\ }\href
  {\doibase 10.1007/JHEP03(2019)034} {\bibfield  {journal} {\bibinfo  {journal}
  {JHEP}\ }\textbf {\bibinfo {volume} {03}},\ \bibinfo {pages} {034} (\bibinfo
  {year} {2019})},\ \Eprint {http://arxiv.org/abs/1809.08251} {arXiv:1809.08251
  [hep-ph]} \BibitemShut {NoStop}%
\bibitem [{\citenamefont {Aristizabal~Sierra}\ \emph
  {et~al.}(2014)\citenamefont {Aristizabal~Sierra}, \citenamefont {Tortola},
  \citenamefont {Valle},\ and\ \citenamefont
  {Vicente}}]{AristizabalSierra:2014uzi}%
  \BibitemOpen
  \bibfield  {author} {\bibinfo {author} {\bibfnamefont {D.}~\bibnamefont
  {Aristizabal~Sierra}}, \bibinfo {author} {\bibfnamefont {M.}~\bibnamefont
  {Tortola}}, \bibinfo {author} {\bibfnamefont {J.~W.~F.}\ \bibnamefont
  {Valle}}, \ and\ \bibinfo {author} {\bibfnamefont {A.}~\bibnamefont
  {Vicente}},\ }\href {\doibase 10.1088/1475-7516/2014/07/052} {\bibfield
  {journal} {\bibinfo  {journal} {JCAP}\ }\textbf {\bibinfo {volume} {07}},\
  \bibinfo {pages} {052} (\bibinfo {year} {2014})},\ \Eprint
  {http://arxiv.org/abs/1405.4706} {arXiv:1405.4706 [hep-ph]} \BibitemShut
  {NoStop}%
\end{thebibliography}%

\end{document}